\pdfoutput=1

\documentclass[12pt,a4paper]{article}

\usepackage{ifthen} 
\newboolean{pdflatex}
\setboolean{pdflatex}{true} 

\newboolean{articletitles}
\setboolean{articletitles}{true} 

\newboolean{uprightparticles}
\setboolean{uprightparticles}{false} 

\newboolean{inbibliography}
\setboolean{inbibliography}{false} 

\usepackage{caption}

\newcommand{\mfig}{Fig.}
\newcommand{\sect}{Sec.}
\newcommand{\sects}{Secs.}
\newcommand{\mfigs}{Figs.}

\usepackage{microtype}
\usepackage{xspace} 

\usepackage{graphicx}  
\usepackage{lineno}
\usepackage{color}
\usepackage{colortbl}
\graphicspath{{./figs/}} 

\usepackage{amsmath} 
\usepackage{amssymb}
\usepackage{amsfonts}
\usepackage{upgreek} 

\newcommand*\patchAmsMathEnvironmentForLineno[1]{%
\expandafter\let\csname old#1\expandafter\endcsname\csname #1\endcsname
\expandafter\let\csname oldend#1\expandafter\endcsname\csname
end#1\endcsname
 \renewenvironment{#1}%
   {\linenomath\csname old#1\endcsname}%
   {\csname oldend#1\endcsname\endlinenomath}%
}
\newcommand*\patchBothAmsMathEnvironmentsForLineno[1]{%
  \patchAmsMathEnvironmentForLineno{#1}%
  \patchAmsMathEnvironmentForLineno{#1*}%
}
\AtBeginDocument{%
\patchBothAmsMathEnvironmentsForLineno{equation}%
\patchBothAmsMathEnvironmentsForLineno{align}%
\patchBothAmsMathEnvironmentsForLineno{flalign}%
\patchBothAmsMathEnvironmentsForLineno{alignat}%
\patchBothAmsMathEnvironmentsForLineno{gather}%
\patchBothAmsMathEnvironmentsForLineno{multline}%
}

\def\lhcb    {\mbox{LHCb}\xspace}
\def\CP                {\ensuremath{C\!P}\xspace}

\def\PB      {\ensuremath{B}\xspace}
\def\Bbar    {\ensuremath{\kern 0.18em\overline{\kern -0.18em \PB}{}}\xspace}
\def\Ps      {\ensuremath{s}\xspace}
\def\squark    {\ensuremath{\Ps}\xspace}
\def\Bsb     {\ensuremath{\Bbar^0_\squark}\xspace}

\def\mum  {\ensuremath{\,\upmu\rm m}\xspace}

\def\gauss      {\mbox{\textsc{Gauss}}\xspace}
\def\pythia     {\mbox{\textsc{Pythia}}\xspace}
\def\evtgen     {\mbox{\textsc{EvtGen}}\xspace}
\def\photos     {\mbox{\textsc{Photos}}\xspace}
\def\geant      {\mbox{\textsc{Geant4}}\xspace}

\def\BF         {{\ensuremath{\cal B}\xspace}}

\def\roofit     {\mbox{\textsc{RooFit}}\xspace}
\def\root       {\mbox{\textsc{Root}}\xspace}

\def\B       {\ensuremath{\PB}\xspace}

\usepackage{cite} 
\usepackage{mciteplus}

\usepackage{hyperref}    
\usepackage[all]{hypcap} 

\newboolean{isepjc}
\setboolean{isepjc}{false}

\begin{document}

\renewcommand{\thefootnote}{\fnsymbol{footnote}}
\setcounter{footnote}{1}


\begin{titlepage}
\pagenumbering{roman}

\vspace*{-1.5cm}
\centerline{\large EUROPEAN ORGANIZATION FOR NUCLEAR RESEARCH (CERN)}
\vspace*{1.5cm}
\hspace*{-0.5cm}
\begin{tabular*}{\linewidth}{lc@{\extracolsep{\fill}}r}
\ifthenelse{\boolean{pdflatex}}
{\vspace*{-2.7cm}\mbox{\!\!\!\includegraphics[width=.14\textwidth]{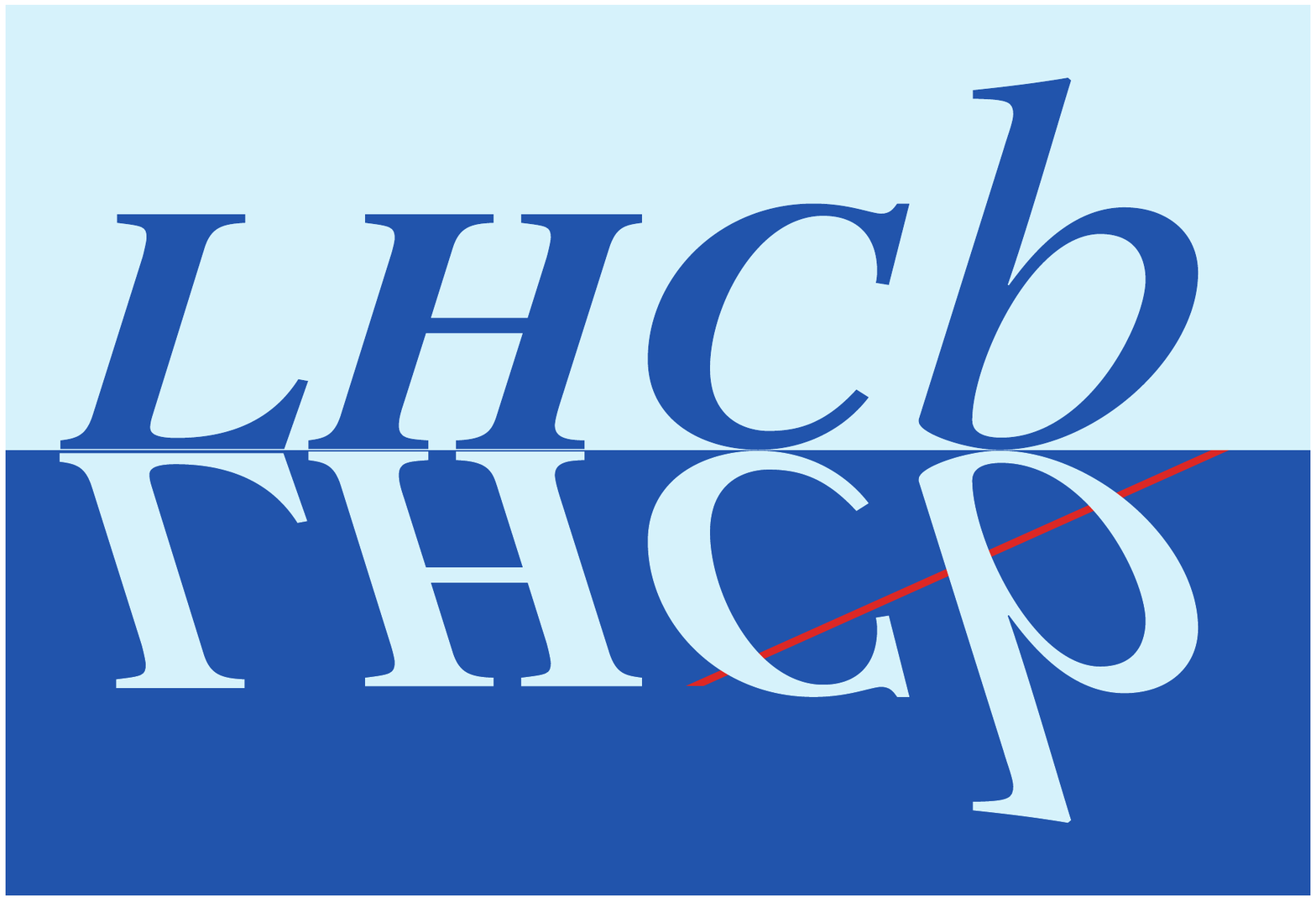}} & &}%
{\vspace*{-1.2cm}\mbox{\!\!\!\includegraphics[width=.12\textwidth]{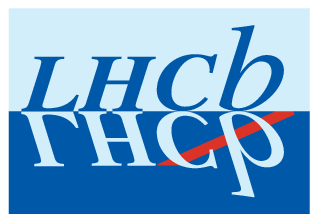}} & &}%
\\
 & & CERN-PH-EP-2013-143 \\  
 & & LHCb-PAPER-2013-036 \\  
 & & 11$^{\textrm{th}}$ October 2013\\
 & & \\

\end{tabular*}

\vspace*{0.5cm}

{\bf\boldmath\huge
\begin{center}
  Observation of \boldmath$B^0_s$-\Bsb mixing and\\ \vspace{2mm} \boldmath$_{}$measurement of mixing frequencies\\ \vspace{2mm} using semileptonic \boldmath$B$ decays
\end{center}
}

\vspace*{0.5cm}

\begin{center}
The LHCb collaboration\footnote{Authors are listed on the following pages.}
\end{center}

\vspace*{1.0cm}

\begin{abstract}
  \noindent
  The $B^0_s$ and $B^0$ mixing frequencies, $\Delta m_s$ and $\Delta m_d$, are measured using a data sample corresponding to an integrated luminosity of 1.0\,fb$^{-1}$ collected by the LHCb experiment in $pp$ collisions at a centre of mass energy of $7$\,TeV during 2011. Around 1.8$\times10^{6}$ candidate events are selected of the type $B^0_{(s)} \to D^-_{(s)} \mu^+$\,($+$\,anything), where about half are from peaking and combinatorial backgrounds. To determine the $B$ decay times, a correction is required for the momentum carried by missing particles, which is performed using a simulation-based statistical method. Associated production of muons or mesons allows us to tag the initial-state flavour and so to resolve oscillations due to mixing. We obtain
\begin{align}
&\Delta m_s = ( 17.93  \pm  0.22\,\textrm{(stat)} \pm  0.15\,\textrm{(syst)}) \,\textrm{ps}^{-1} \nonumber ,\\
&\Delta m_d = ( 0.503  \pm  0.011\,\textrm{(stat)} \pm  0.013\,\textrm{(syst)}) \,\textrm{ps}^{-1}\nonumber .
\end{align}

The hypothesis of no oscillations is rejected by the equivalent of 5.8 standard deviations for $B^0_s$ and 13.0 standard deviations for $B^0$. This is the first observation of $B^0_s$ mixing to be made using only semileptonic decays.

\end{abstract}

\vspace*{1.0cm}

\begin{center}
  To be published in Eur.~Phys.~J.~C
\end{center}


{\footnotesize
\centerline{\copyright~CERN on behalf of the \lhcb collaboration, license \href{http://creativecommons.org/licenses/by/3.0/}{CC-BY-3.0}.}}
\vspace*{2mm}

\end{titlepage}



\centerline{\large\bf LHCb collaboration}
\begin{flushleft}
\small
R.~Aaij$^{40}$, 
B.~Adeva$^{36}$, 
M.~Adinolfi$^{45}$, 
C.~Adrover$^{6}$, 
A.~Affolder$^{51}$, 
Z.~Ajaltouni$^{5}$, 
J.~Albrecht$^{9}$, 
F.~Alessio$^{37}$, 
M.~Alexander$^{50}$, 
S.~Ali$^{40}$
% ...

{\footnotesize \it
% ...
$ ^{5}$Clermont Universit\'{e}, Universit\'{e} Blaise Pascal, CNRS/IN2P3, LPC, Clermont-Ferrand, France\\
$ ^{6}$CPPM, Aix-Marseille Universit\'{e}, CNRS/IN2P3, Marseille, France\\
% ...
$ ^{9}$Fakult\"{a}t Physik, Technische Universit\"{a}t Dortmund, Dortmund, Germany\\
% ...
$ ^{36}$Universidad de Santiago de Compostela, Santiago de Compostela, Spain\\
$ ^{37}$European Organization for Nuclear Research (CERN), Geneva, Switzerland\\
% ...
$ ^{40}$Nikhef National Institute for Subatomic Physics, Amsterdam, The Netherlands\\
% ...
$ ^{45}$H.H. Wills Physics Laboratory, University of Bristol, Bristol, United Kingdom\\
% ...
$ ^{50}$School of Physics and Astronomy, University of Glasgow, Glasgow, United Kingdom\\
$ ^{51}$Oliver Lodge Laboratory, University of Liverpool, Liverpool, United Kingdom\\
% ...
}
\end{flushleft}

\clearpage


\renewcommand{\thefootnote}{\arabic{footnote}}
\setcounter{footnote}{0}


\pagestyle{plain} 
\setcounter{page}{1}
\pagenumbering{arabic}

\section{Introduction}
\label{Section:Introduction}

$B_s^0$ and $B^0$ mesons propagate as superpositions of particle and antiparticle flavour states. For a flavour-specific decay process\footnote{In this paper, charge conjugate modes are always implied.} such as $B^0~\to~D^-\mu^+\nu$, particle-antiparticle mixing lends a sinusoidal component to the decay rates~\cite{PDG2012,Schneider:B0mixB:2006}. To measure mixing, the flavour state of the $B$ meson must be observed to change, which requires knowledge of the state from at least two points in time. The experimentally accessible times to determine the flavour are at production and decay.
Neglecting \CP violation in mixing, the decay rate $N$ at a proper decay time $t$ simplifies to\begin{equation}
N_\pm(t)= N(0)\, \frac{e^{-\Gamma t}}{2}\left[\cosh{(\Delta\Gamma_{\,} t/2)}\pm\cos{(\Delta m_{\,} t)}\right]\,, \label{Eqn:Intro:UnTagged}
\end{equation}
where $\Delta\Gamma$ and $\Delta m$ are the width and mass differences\footnote{The mass difference is measured here as an angular frequency, in units of inverse time.} of the two mass eigenstates, and $\Gamma$ is the average decay width~\cite{Schneider:B0mixB:2006}.
The positive sign applies when the $B$ meson decays with the same flavour as its production and the negative sign when the particle decays with opposite flavour to its production, later referred to as ``even'' and ``odd''. In this study, a sample of semileptonic decays obtained with the LHCb detector is used to measure the mixing frequencies $\Delta m_s$ and $\Delta m_d$ for the $B^0_s$ and $B^0$ systems. These quantities have previously been measured to high precision, usually in the combination of several channels, relying heavily on hadronic decay modes (see for example Refs.~\cite{Albrecht:1987dr, Abulencia:2006ze} and our recent results, Refs.~\cite{LHCb-PAPER-2013-006, LHCb-PAPER-2011-027, LHCB-PAPER-2012-032}). To date no observation of $B_s^0$ mixing has been made using only semileptonic decay channels.

\section{Experimental setup}
\label{Section:LHCb}

The LHCb detector~\cite{Alves:2008zz} is a single-arm forward
spectrometer covering the pseudorapidity range $2<\eta <5$,
designed for the study of particles containing $b$ or $c$
quarks. The detector consists of several dedicated subsystems, organized successively further from the interaction region.
A silicon-strip vertex detector surrounds the $pp$
interaction region and approaches to within 8\,mm of the proton beams. The first of two ring-imaging Cherenkov (RICH) detectors comes next, followed by the remainder of the tracking system, which comprises, in order: a large-area silicon-strip detector; a dipole magnet with a bending power of about
$4{\rm\,Tm}$; and three multilayer tracking stations, each with central silicon-strip detectors and peripheral straw
drift tubes. After this comes the second RICH detector, the calorimeter and the muon stations.

The combined high-precision tracking system provides a momentum measurement with
relative uncertainty that varies from 0.4\,\% at 5\,GeV$c^{-1}$ to 0.6\,\% at 100\,GeV$c^{-1}$,
and impact parameter\footnote{The impact parameter is the distance of closest approach of a track to a primary interaction vertex.} resolution of 20\,\mum for
tracks with high transverse momentum. By combining information from the two RICH detectors~\cite{LHCb-DP-2012-003} charged hadrons can be identified across a wide range in momentum, around 2 to 150\,GeV$c^{-1}$. The calorimeter system consists of
scintillating-pad and preshower detectors, an electromagnetic
calorimeter and a hadronic calorimeter, allowing identification of photon, electron and
hadron candidates. Muons that pass through the calorimeters are detected using a system of alternating layers of iron and multiwire
proportional chambers~\cite{LHCb-DP-2012-002}. Triggering of events is performed in two stages~\cite{LHCb-DP-2012-004}: a
hardware stage, based on information from the calorimeter and muon
systems, followed by a software stage, which performs full event reconstruction.


\section{Data selection and reconstruction}
\label{Section:Data}

The LHCb dataset used in this analysis corresponds to an integrated luminosity of 1.0\,fb$^{-1}$ collected in $pp$ collisions at a centre of mass energy of $7$\,TeV during the 2011 physics run at the LHC. Where simulation is required,
\pythia~6.4~\cite{Sjostrand:2006za} is used, with a specific LHCb
configuration~\cite{LHCb-PROC-2010-056}.  Decays of hadronic particles
are described by \evtgen~\cite{Lange:2001uf}, in which final-state
radiation is generated using \photos~\cite{Golonka:2005pn}. The
interaction of the generated particles with the detector and the detector
response are implemented using the \geant
toolkit~\cite{Allison:2006ve, *Agostinelli:2002hh} as described in
Ref.~\cite{LHCb-PROC-2011-006}. Input to \evtgen is taken from the best knowledge of branching fractions (\BF) and form factors at the time of the simulation~\cite{PDG2012}. The same reconstruction and selection is applied on simulated and detector data.

A sample of events is selected in which a $D_{(s)}^+ \to K^{+}K^{-}\pi^+$ candidate forms a vertex with a muon candidate. A cut-based selection is applied to enhance the fraction of real $D^+_{(s)}$ mesons in this sample that arise from $B^0_{(s)}$ semileptonic decays.
Vertex and track reconstruction qualities, momenta, invariant masses, flight distances and particle identification (PID) variables are used. The selection was initially optimized on simulated data to maximize the signal significance, $S/\sqrt{(S+B)}$, where $S$~($B$) denotes the number of selected signal (background) candidates. The most important cuts for this analysis are those on the PID and invariant masses. Combined information from the RICH detectors, muon stations, calorimeters and tracking allows us to place stringent requirements on a log-likelihood based PID parameter for each final-state particle separately, ensuring at least 99\,\% purity in the muon sample, and suppressing peaking backgrounds such as $D^+ \to K^-\pi^+\pi^+$ decays, where a pion has been misidentified as a kaon. To allow a simultaneous measurement of $\Delta m_s$ and $\Delta m_d$, a broad mass window for the $K^+K^-\pi^+$ system is used to cover both the $D^+$ and $D_s^+$ masses, $-0.2 < M(K^+K^-\pi^+) - M_0(D^+_s) < 0.1$\,GeV$c^{-2}$, where $M_0(D^+_s)$ is the known mass of the $D^+_s$ meson~\cite{PDG2012}. Decays of the type $D^{\ast}(2010)^{+}\to D^0\pi^+$ are additionally suppressed by requiring that the invariant mass of the two kaons $M(K^+K^-)\,<\,1.84$\,GeV$c^{-2}$, and combinatorial background with slow collinear pions is similarly removed with the mass requirement $M(K^+K^-\pi^+)-M(K^+K^-)-M_0(\pi^+)>15$\,MeV$c^{-2}$.

Simulation studies indicate that the selected sample is dominated by $B_s^0 \to D_s^{-} \mu^{+} (\nu, \pi^0, \gamma)$, $B^0 \to D^{-} \mu^{+} (\nu, \pi^0, \gamma)$ and $B^+ \to D^{-} \mu^{+} (\nu, \pi^+, \gamma)$ decays, where no specific intermediate states are required other than those mentioned, and where at least one neutrino will occur together with any number of the other particles in the parentheses. These additional particles are ignored and so a clear $B$ mass peak cannot be reconstructed. For simplicity, to quantify the measured mass, $M(D\mu)$, within its possible range, we define a ``normalized mass'', $n$, relative to the known masses $(M_0)$ of the $B$, $D$, and $\mu$:
\begin{equation}
  n = \frac{M(D\mu)-M_0(D)-M_0(\mu)}{M_0(B)-M_0(D)-M_0(\mu)}. \label{Equation:n}
\end{equation}
We require $0.24<n<1.0$, where the lower cut mainly removes low-mass combinatorial background candidates. The $K^+K^-\pi^+$ invariant mass distribution and the normalized mass distribution ($n$) of the selected candidates are shown in \mfig~\ref{Figure:AllData}, in which the $D^+_s$ and $D^+$ peaks can clearly be seen over the combinatorial background.

Determination of the initial-state flavour is performed using the standard LHCb flavour-tagging algorithms, which are described in detail elsewhere~\cite{MGrabalosa:Thesis, LHCb-PAPER-2011-027, LHCb-PAPER-2013-006}. These algorithms rely on the reconstruction of particles that were produced in association with, and are flavour-correlated with, the signal $B$-meson. The correlations arise either from fragmentation, which often produces a kaon or pion of specific charge correlated with the signal, or from ``opposite-side'' decays, where the decay products of the partner $b$ quark are reconstructed (e.g.~a muon). A neural network combines tagging decisions for the best tagging power\cite{LHCb-PAPER-2011-027}.

\ifthenelse{\boolean{isepjc}}%
{
\begin{figure*}
  \includegraphics[width=\textwidth,keepaspectratio,]{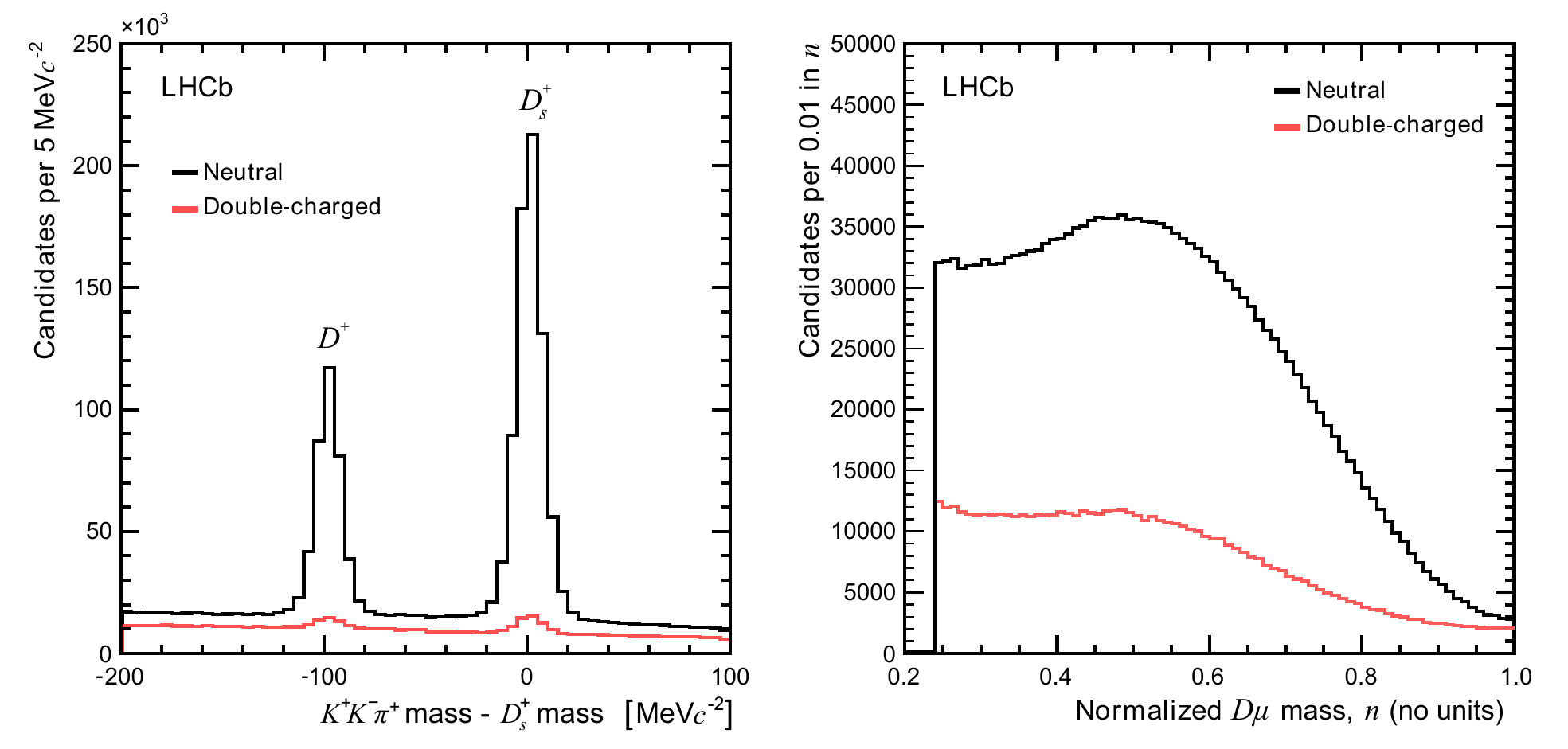}
  \caption{Mass distributions for all selected signal candidates. Left, the $K^+K^-\pi^+$ invariant mass, where the known mass of the $D^+_s$ has been subtracted. Right, the $D\mu$ normalized mass as defined in Eq.~\ref{Equation:n}. Neutral candidates are those of the form $D^{\mp}\mu^\pm$, while double-charged candidates are those of the form $D^{\pm}\mu^\pm$. The double-charged candidates arise from several background sources, most of which are also present in the neutral sample. In the left plot, the neutral sample exhibits much larger $D$ mass peaks, indicative of the large $B$ signal component.}
  \label{Figure:AllData}
\end{figure*}
}
{
\begin{figure}[b]
  \centering
  \includegraphics[width=\textwidth,keepaspectratio,]{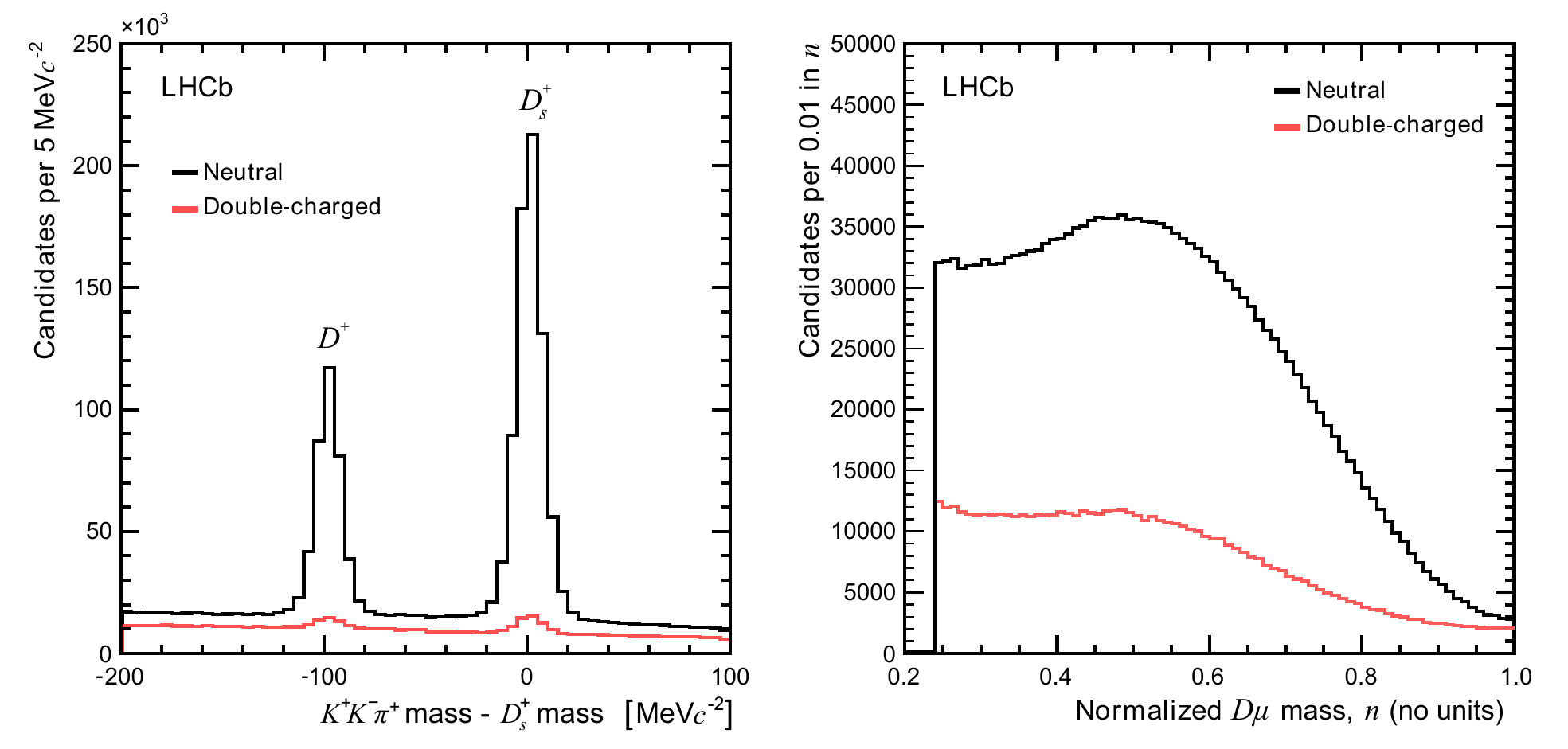}
  \caption{Mass distributions for all selected signal candidates. Left, the $K^+K^-\pi^+$ invariant mass, where the known mass of the $D^+_s$ has been subtracted. Right, the $D\mu$ normalized mass as defined in Eq.~\ref{Equation:n}. Neutral candidates are those of the form $D^{\mp}\mu^\pm$, while double-charged candidates are those of the form $D^{\pm}\mu^\pm$. The double-charged candidates arise from several background sources, most of which are also present in the neutral sample. In the left plot, the neutral sample exhibits much larger $D$ mass peaks, indicative of the large $B$ signal component.}
  \label{Figure:AllData}
\end{figure}
}

A hypothesis is required for the nature of the reconstructed candidate, either $B^0_s$ or $B^0$, in order to choose the tagging algorithms to be applied and to select the appropriate mass with which to calculate $n$. A split around the midpoint between the $D^+_s$ and $D^+$ peaks is used. For the $B^0_s$ hypothesis all available tags are used. For the $B^0$ hypothesis only opposite-side tags are used, to reduce the difference between $B^+$ and $B^0$ tagging performance and thus better constrain the $B^+$ background (see \sects~\ref{Section:Fit}~and~\ref{Section:Systematics}). The flavour-tagged dataset comprises 594,845 selected candidates.

Two techniques are employed to measure the mixing frequencies: (a) multidimensional log-likelihood maximization, simultaneously fitting $\Delta m_s$ and $\Delta m_d$; (b) model-independent Fourier analysis, used as a cross-check, which determines $\Delta m_s$ with good precision, but $\Delta m_d$ with a very poor precision. Both methods use a common determination of the proper decay time and so share a portion of the corresponding systematic effects.


\section{Proper decay-time distributions}
\label{Section:Time}

\ifthenelse{\boolean{isepjc}}%
{
\begin{figure}
  \includegraphics[width=\columnwidth,keepaspectratio,]{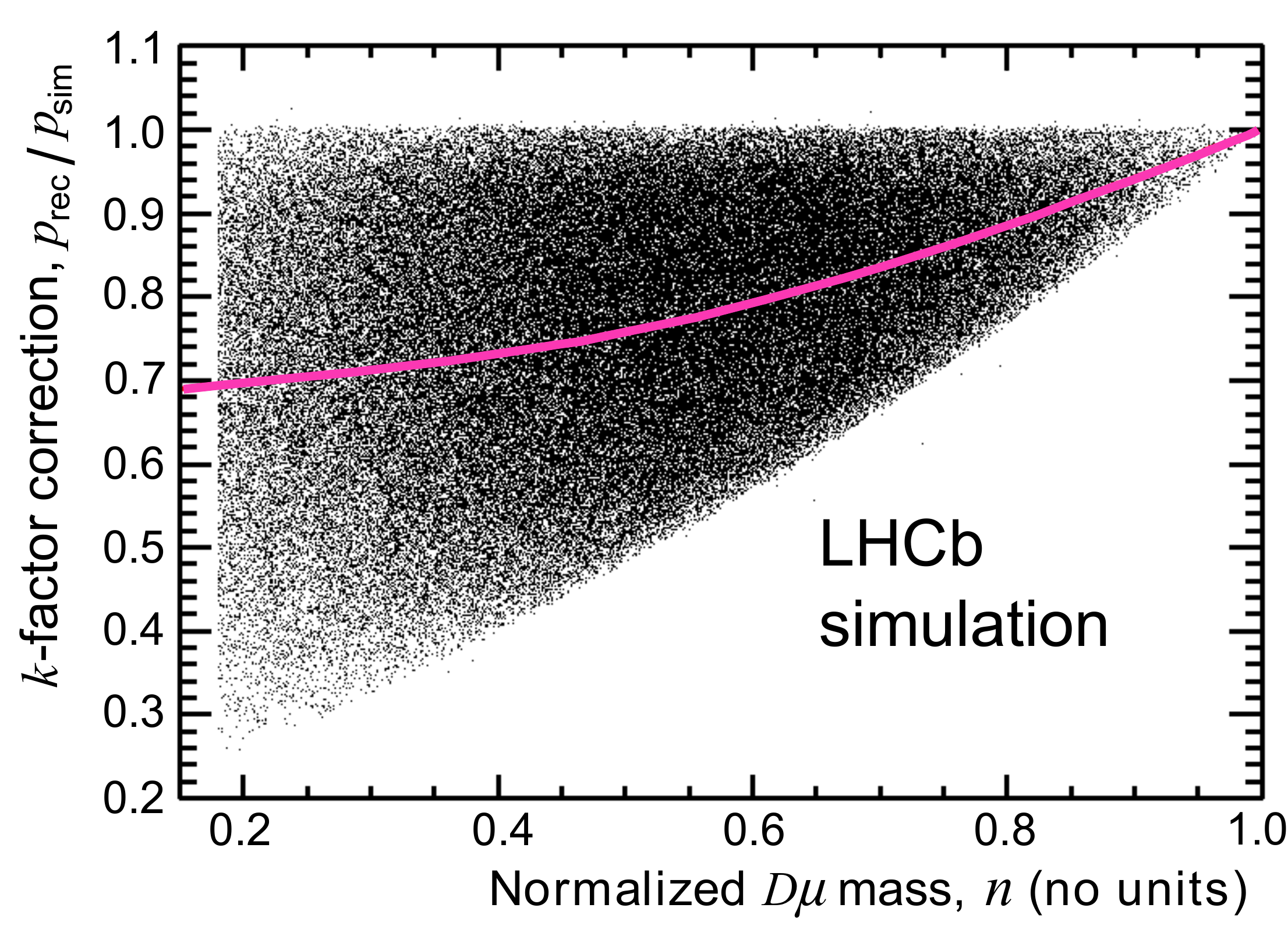}
  \caption{Input to obtain the $k$-factor correction from the fully-simulated $B^0_s$ sample. For each event the ratio of reconstructed to generated momentum, $p_{\textrm{rec}}/p_{\textrm{sim}}$ is plotted against the normalized $D\mu$ mass ($n$ in Eq.~\ref{Equation:n}). The curve shows a fourth-order polynomial resulting from a fit to the mean of the distribution (in bins of $n$).}
  \label{Figure:K:BsFits}
\end{figure}
}
{
\begin{figure}[bp]
  \centering
  \includegraphics[width=0.60\textwidth,keepaspectratio,]{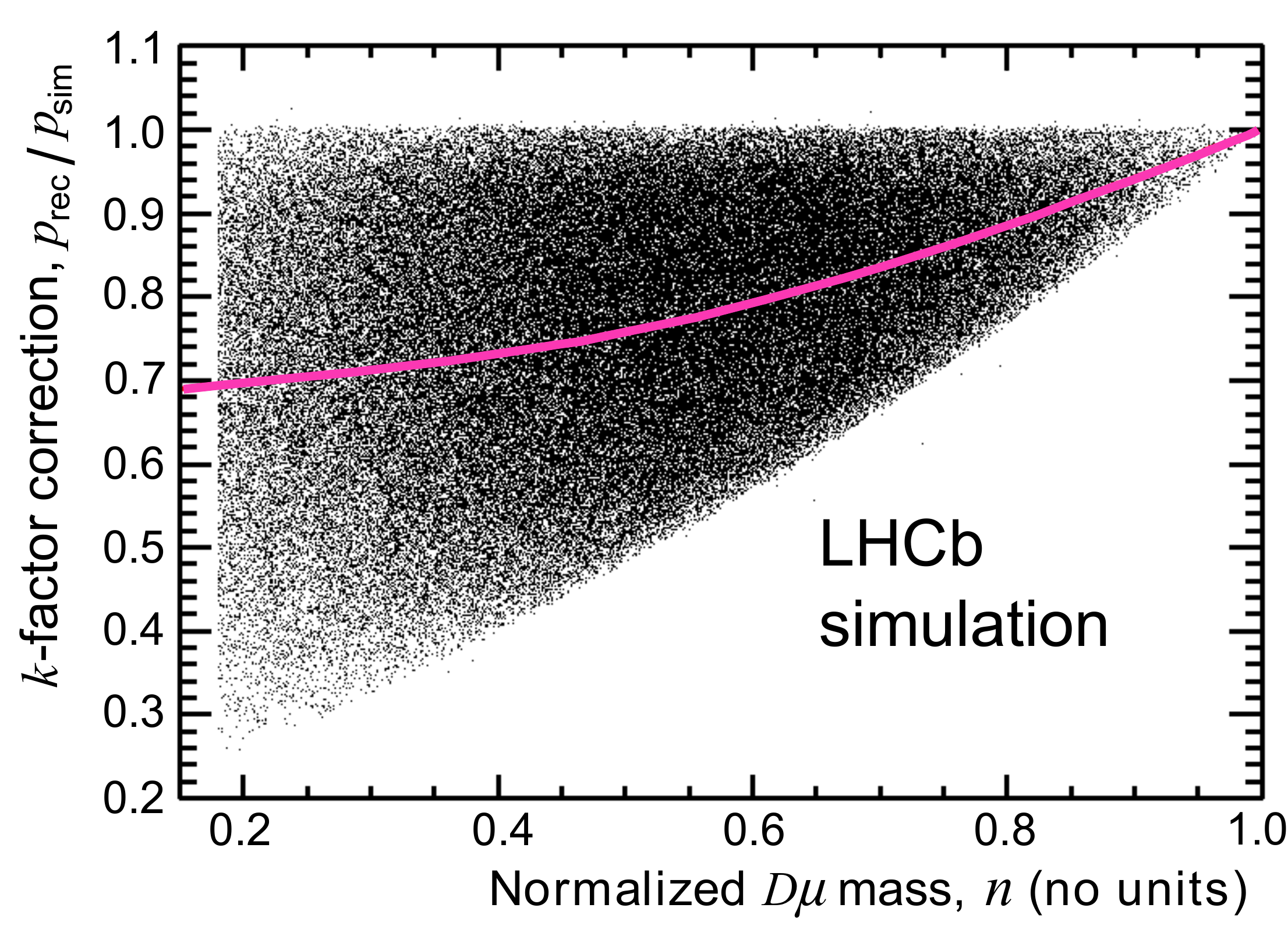}
  \caption{Input to obtain the $k$-factor correction from the fully-simulated $B^0_s$ sample. For each event the ratio of reconstructed to generated momentum, $p_{\textrm{rec}}/p_{\textrm{sim}}$ is plotted against the normalized $D\mu$ mass ($n$ in Eq.~\ref{Equation:n}). The curve shows a fourth-order polynomial resulting from a fit to the mean of the distribution (in bins of $n$).}
  \label{Figure:K:BsFits}
\end{figure}
}

\ifthenelse{\boolean{isepjc}}%
{
\begin{figure*}
  \begin{center}
    \includegraphics[width=\textwidth,keepaspectratio,]{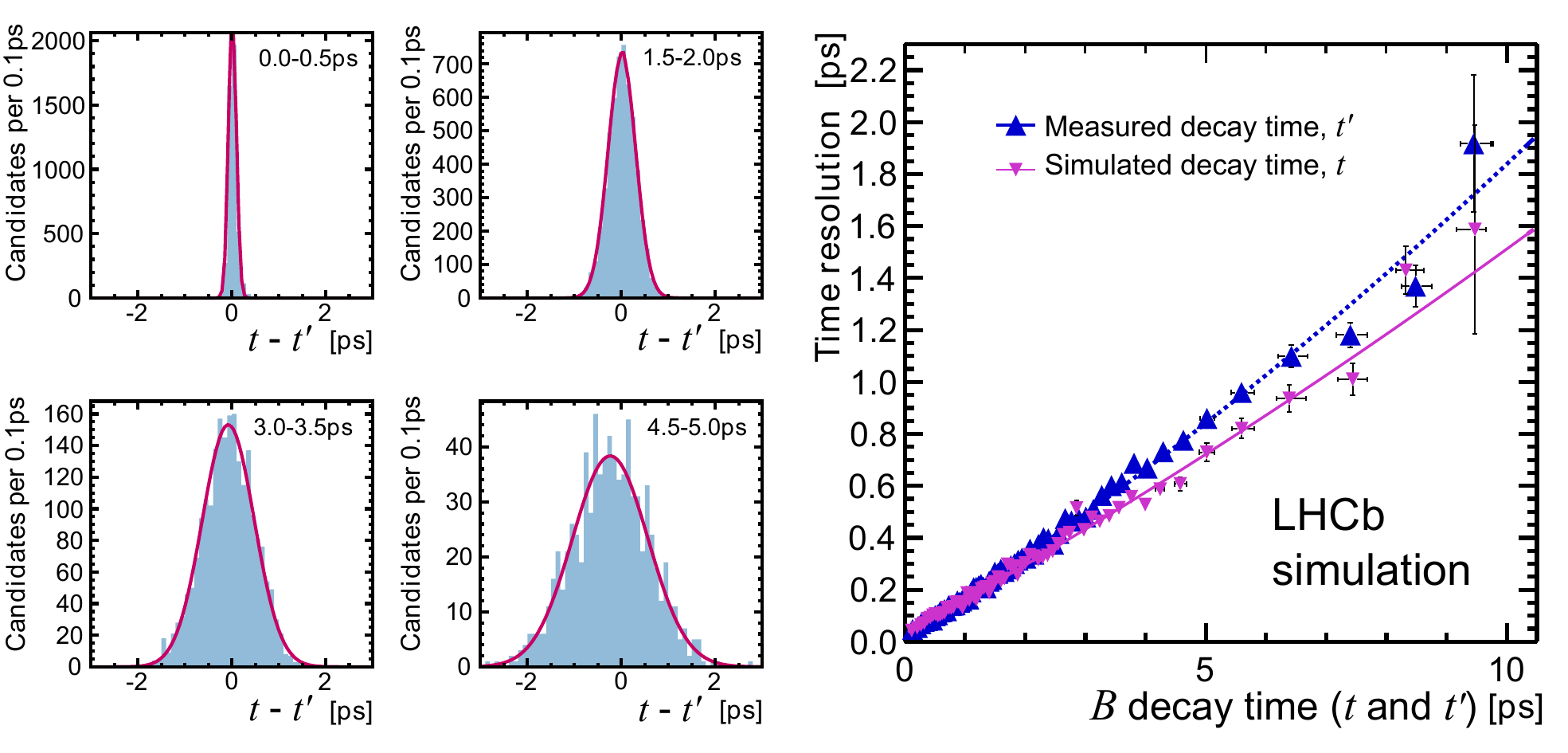}
\end{center}
    \caption{Illustration of the decay time resolution obtained from a fully simulated $B^0$ signal sample. The left plots demonstrate the Gaussian fits (solid lines) using the full LHCb simulated data (filled), to determine the decay time resolution. Each measured (reconstructed and corrected) time, $t'$, is compared to the corresponding simulated decay time, $t$. The results are shown for several bins of $t'$. The dependence on decay time of the mean (bias, $\mu$) and width (standard deviation, $\sigma$) can be fitted with a quadratic or cubic function of either $t$ or $t'$. The right hand plot shows a quadratic fit to the widths. }\label{Figure:Toys:ResolutionBd}
\end{figure*}
}{}

To obtain the $B$-meson decay times, a correction is applied for the momentum lost due to missing particles, using a $k$-factor method as employed in many previous measurements (see, for example, Refs.~\cite{CDF:Dms:Thesis:2006} and \cite{D0:Dms:Thesis:2008}).
The $k$-factor \cite{TBird:Thesis:2011} is a simulation-based statistical correction, where the average missing momentum in a simulated sample is used to correct the reconstructed momentum as a function of the reconstructed $D\mu$ mass (as shown in \mfig~\ref{Figure:K:BsFits}). In this study we use a fourth-order polynomial to parameterize $k$ as a function of the normalized $D\mu$ mass ($n$ from Eq.~\ref{Equation:n}), which allows us to use the same correction for $B^0_s$ and $B^0$. With this approach, both $\Delta m_s$ and $\Delta m_d$ exhibit residual biases of around $1$\,\%; these biases are known to good precision from the full simulation and are corrected in the final results.

The experimental resolution of the proper decay time ($t$) reduces the visibility of the oscillations, smearing Eq.~\ref{Eqn:Intro:UnTagged} with a resolution function $R(t,t'-t)$, where $t$ is the true decay time and $t'$ is the measured value. The limited performance of the tagging also reduces the visibility of the oscillations. Our selection requirements include variables that are correlated with the decay time, leading to a time-dependent efficiency function, $\varepsilon(t')$. Thus Eq.~\ref{Eqn:Intro:UnTagged} becomes\ifthenelse{\boolean{isepjc}}%
{
\begin{align}
N_{\pm}(t')= N(0)\,\eta\, \frac{e^{-\Gamma t}}{2}\Bigl[& \cosh{(\Delta\Gamma \, t/2)} \nonumber\\
                                                       & \pm(1-2\omega)\cos{(\Delta m \, t)} \Bigr] \nonumber\\
                                                  & \otimes R(t,t'-t) \times \varepsilon(t'), \label{Eqn:Intro:TaggedRes}
\end{align}}{
\begin{align}
N_{\pm}(t')= N(0)\,\eta\, \frac{e^{-\Gamma t}}{2}\left[\cosh{(\Delta\Gamma \, t/2)}\pm(1-2\omega)\cos{(\Delta m \, t)}\right] \otimes R(t,t'-t) \times \varepsilon(t'), \label{Eqn:Intro:TaggedRes}
\end{align}}
where $\eta$ is the tagging efficiency and $\omega$ is the mistag probability (the fraction of tags that assign the wrong flavour). We parameterize the time-dependent efficiency with an empirical ``acceptance'' function. Specifically Gaussian functions are used as motivated by data and full simulation studies~\cite{TBird:Thesis:2011}, $\varepsilon(t')= 1-f\,G(t';\mu_0,\sigma_1)-(1-f)\,G(t';\mu_0,\sigma_2)$, where $G$ is the Gaussian function and the parameters are determined from fits to the data (typical values are $\sigma_{1,2}<1$\,ps and $\mu_0 \approx 0.01$\,ps).

\ifthenelse{\boolean{isepjc}}%
{}
{
\begin{figure}[b]
  \begin{center}
    \includegraphics[width=\textwidth,keepaspectratio,]{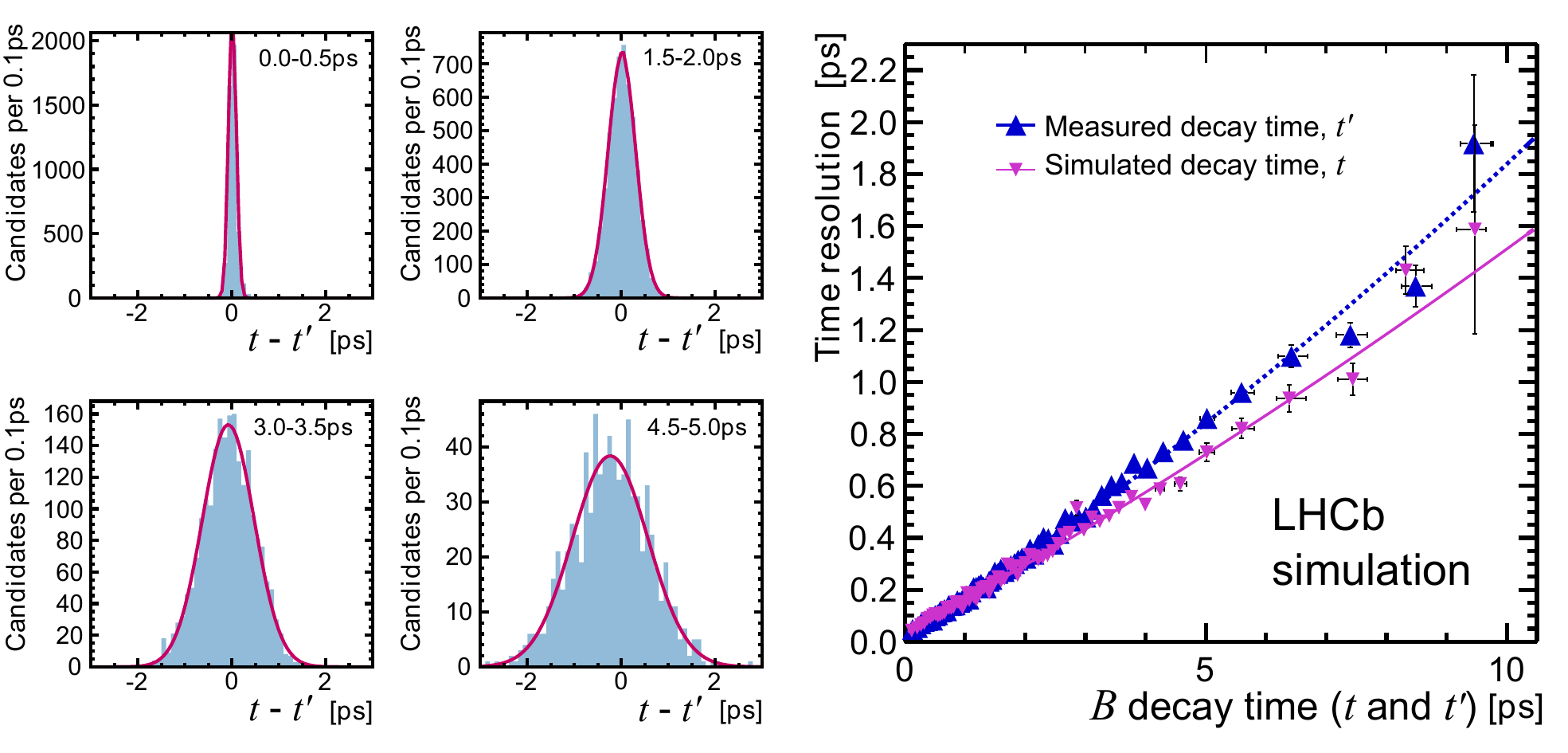}
\end{center}
    \caption{Illustration of the decay time resolution obtained from a fully simulated $B^0$ signal sample. The left plots demonstrate the Gaussian fits (solid lines) using the full LHCb simulated data (filled), to determine the decay time resolution. Each measured (reconstructed and corrected) time, $t'$, is compared to the corresponding simulated decay time, $t$. The results are shown for several bins of $t'$. The dependence on decay time of the mean (bias, $\mu$) and width (standard deviation, $\sigma$) can be fitted with a quadratic or cubic function of either $t$ or $t'$. The right hand plot shows a quadratic fit to the widths. }\label{Figure:Toys:ResolutionBd}
\end{figure}
}

The $k$-factor is a relative correction for the average missing momentum at a given value of $n$; as shown in \mfig~\ref{Figure:K:BsFits}, the range of missing momenta is broad and varies from about 70\,\% at $n = 0.2$ to zero at $n = 1$.  This large relative uncertainty on the corrected momentum ($p'$) dominates the decay time resolution, i.e.~$\sigma(t')/t' \approx \sigma(p')/p'$. Hence $\sigma(t')$ is approximately proportional to $t'$ (as seen in \mfig~\ref{Figure:Toys:ResolutionBd}) and the decay time resolution worsens as decay time increases. This dependence is determined and parameterized from the full simulation. We may choose between a parameterization in terms of either the generated (``true'') decay time, using a numerical convolution, or in terms of the measured decay time, using analytical methods; the latter is the default approach. The resolution dependence is well-fitted with second or third order polynomials.

\section{Multivariate fits to the data}
\label{Section:Fit}

\ifthenelse{\boolean{isepjc}}%
{
\begin{figure}
  \includegraphics[width=\columnwidth,keepaspectratio,]{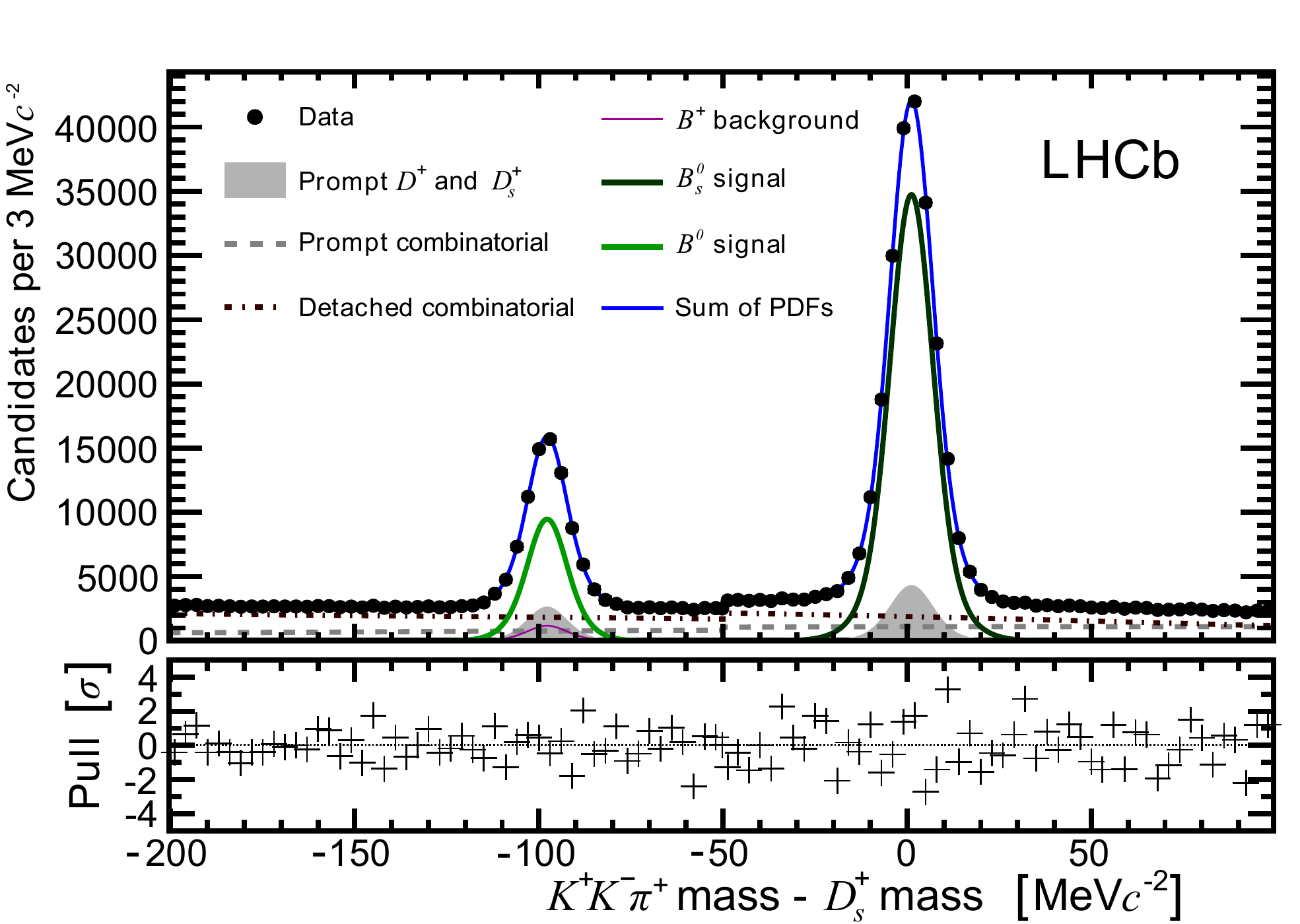}
 \caption{Distribution of measured $K^+K^-\pi^+$ mass, where the known mass of the $D^+_s$ has been subtracted. Black points show the data, and the various lines overlay the result of the fit. The small step at $-50$\,MeV$c^{-2}$ is the result of differences in tagging efficiency for the $B^0_s$ and $B^0$ hypotheses.}
  \label{Figure:model:mass}
\end{figure}
}
{
\begin{figure}[b]
\begin{center}
  \includegraphics[width=0.9\textwidth,keepaspectratio,]{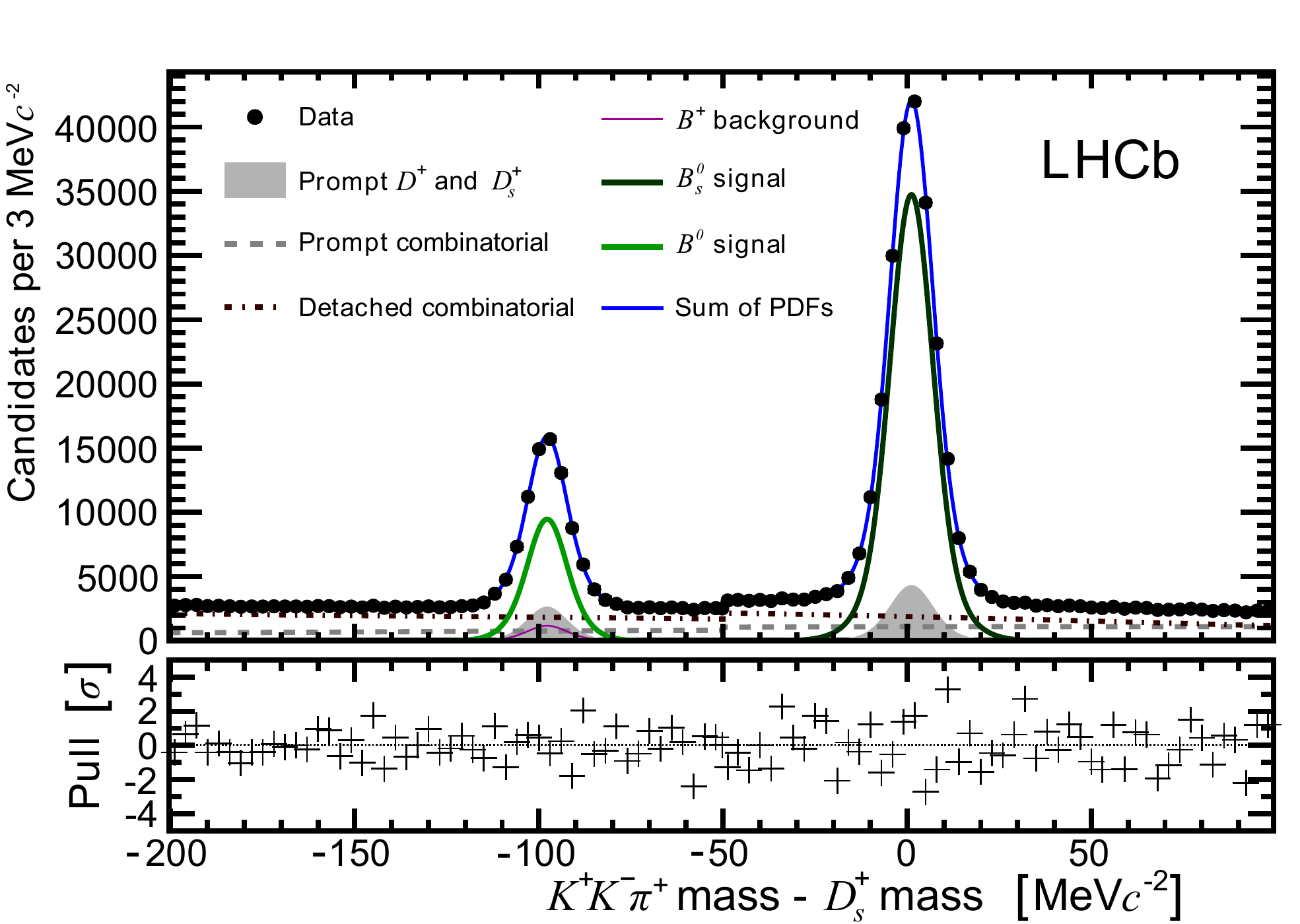}
 \end{center}
 \caption{Distribution of measured $K^+K^-\pi^+$ mass, where the known mass of the $D^+_s$ has been subtracted. Black points show the data, and the various lines overlay the result of the fit. The small step at $-50$\,MeV$c^{-2}$ is the result of differences in tagging efficiency for the $B^0_s$ and $B^0$ hypotheses.}
  \label{Figure:model:mass}
\end{figure}
}

A binned, multidimensional, log-likelihood fit to the data is made, using the \root and embedded \roofit fitting frameworks~\cite{ROOT2,Verkerke:2003irmod}. In order to improve the resolution on the fitted value of $\Delta m_s$, the sample is divided into two subsamples about normalized mass $n=0.56$ (with this value determined using fast-simulation ``pseudo-experiment'' studies), and the two subsamples are fitted simultaneously as described below. There are 101,000 bins over the $K^+K^-\pi^+$ mass, the measured decay time ($t'$), the normalized mass ($n<0.56$ and $n>0.56$), and the tagging result (even and odd). Seven categories of signal and background are assigned component probability density functions (PDFs) whose fractions and shape parameters are left free in the fits to the data. The backgrounds are categorized in terms of their shapes in the mass and decay-time observables. Using the $M(K^+K^-\pi^+)$ distribution we separate out peaking $D_{(s)}^+$ components from combinatorial background components. Each of these categories can be further divided into two based on their decay-time shape. We use the term ``prompt'' to describe fake candidates containing particles exclusively produced in the primary $pp$ interaction, and the term ``detached'' for candidates that contain at least one daughter of a secondary decay and which therefore tend to exhibit a significantly larger lifetime.  Candidates for the signal $B$-decays of interest must be both detached and peaking. The signal-like decays are usually grouped together in the fit; however, we separate the specific background contribution of $B^+$ within the $D^+$ peak and fit that directly. These components are shown in together in \mfig~\ref{Figure:model:mass}~and separately in different $M(K^+K^-\pi^+)$ regions in \mfigs~\ref{Figure:model:tauB}~and~\ref{Figure:model:tau}.
\ifthenelse{\boolean{isepjc}}%
{
}{
\begin{figure}[b]
  \centering
  \includegraphics[width=0.48\textwidth,keepaspectratio,clip=true,trim=0.4cm 0cm 0cm 1cm]{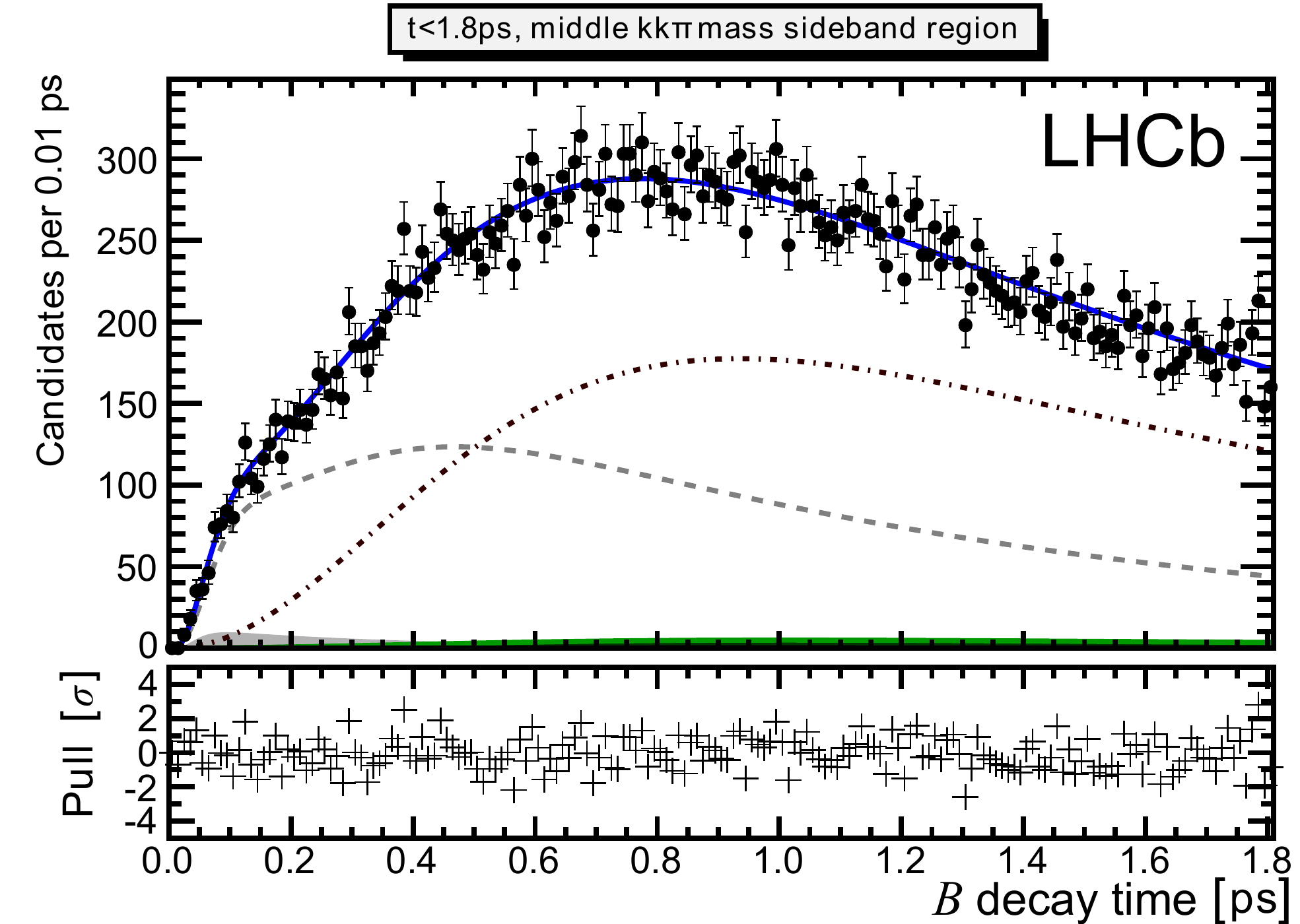}\hspace{1mm}
  \includegraphics[width=0.48\textwidth,keepaspectratio,clip=true,trim=0.4cm 0cm 0cm 1cm]{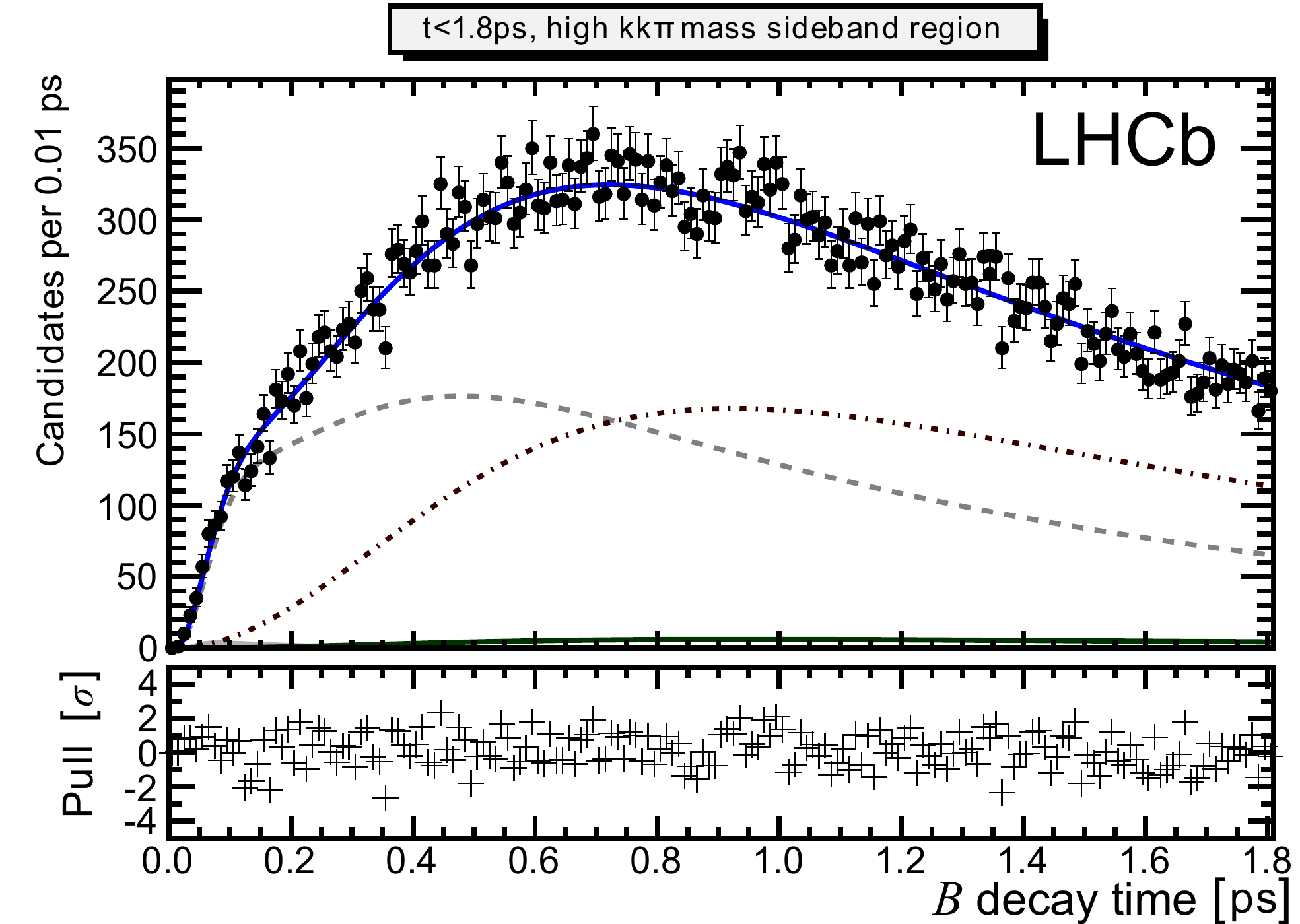}\\ \vspace{1mm}
  \includegraphics[width=0.478\textwidth,keepaspectratio,clip=true,trim=0.4cm 0cm 0cm 1cm]{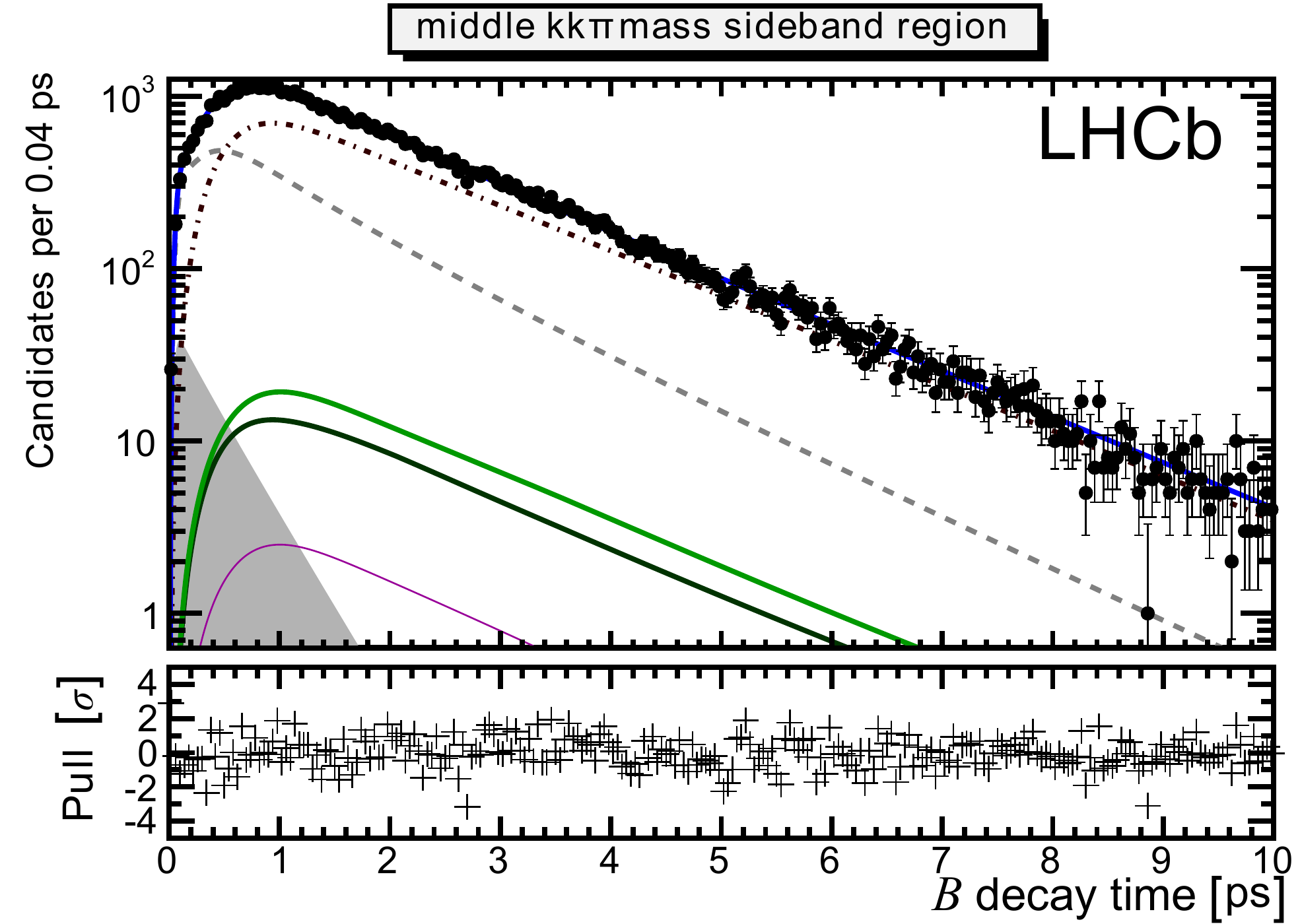}\hspace{0.2mm} 
  \includegraphics[width=0.478\textwidth,keepaspectratio,clip=true,trim=0.4cm 0cm 0cm 1cm]{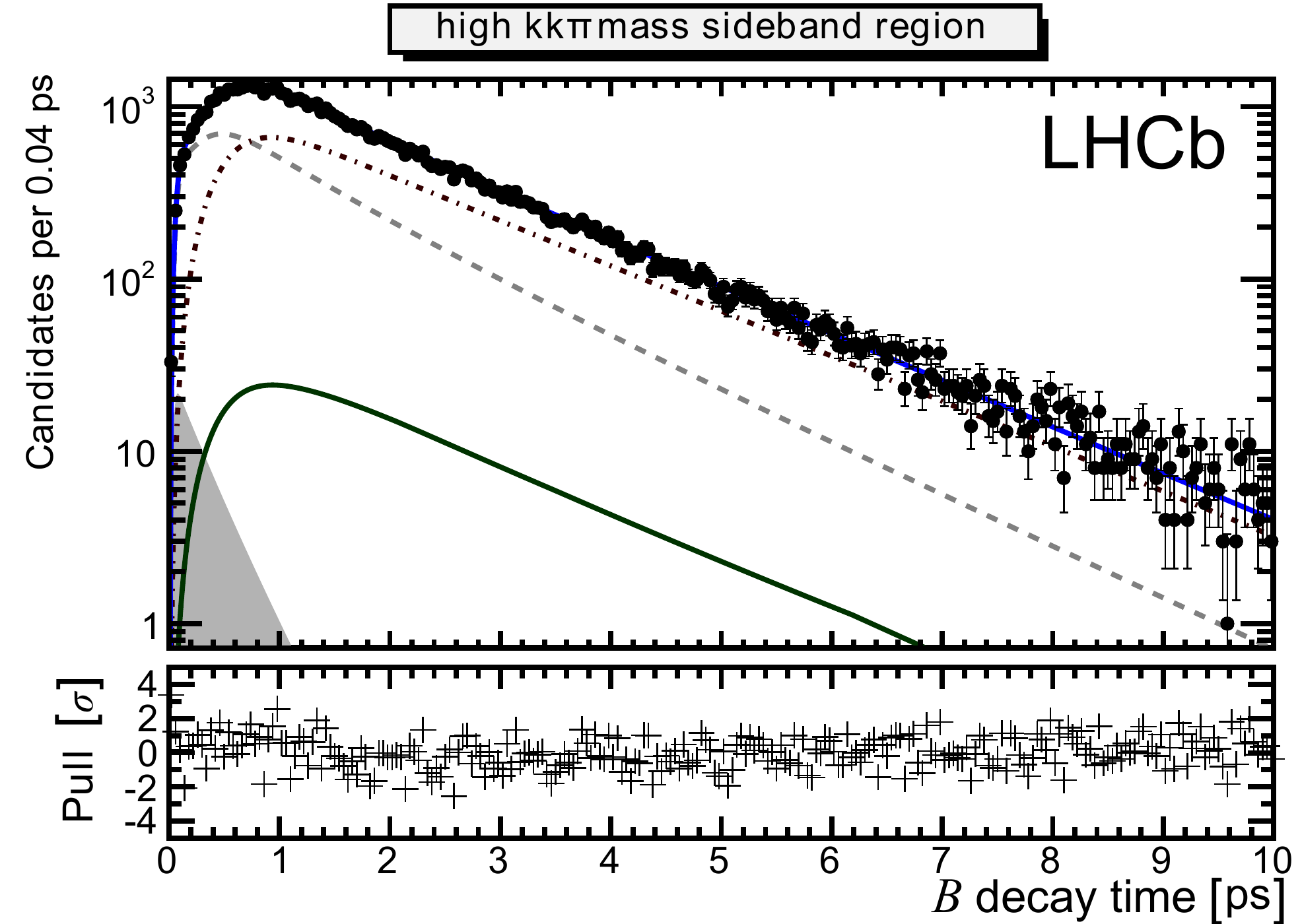} 
  \caption{Measured $B$ decay-time distribution, overlaid with projections of the fit, for background-only regions. Top left: a region between the two signal peaks, $-80$~to~$-20$\,MeV$c^{-2}$ (with respect to the known mass of the $D_s^+$), showing only low decay times. Top right: a region to the right of the signal peaks $20$~to~$100$\,MeV$c^{-2}$, showing only low decay times. Bottom row: the same on an extended decay-time scale and logarithmic. The legend is the same as in \mfig~\ref{Figure:model:mass}.}
  \label{Figure:model:tauB}
\end{figure}
} Each mass PDF is a Gaussian function or a Chebychev polynomial (\mfig~\ref{Figure:model:mass}), and each background decay-time PDF is a simple exponential with an appropriate acceptance function as previously described (\mfig~\ref{Figure:model:tau}). For the signal decay-time shape we use the model described in Eq.~\ref{Eqn:Intro:TaggedRes}, with one instance for each peak. The majority of our sensitivity arises from the mixing asymmetry, whose time-dependent fit in the signal regions is shown in \mfig~\ref{Figure:signal-asym}. Any odd/even asymmetry is assumed to be constant as a function of time for prompt backgrounds and for backgrounds that are known not to mix ($B^+, \Lambda_b$, etc.). Generic detached backgrounds are allowed to have a time-dependent asymmetry varying as an arbitrary quadratic polynomial.

\ifthenelse{\boolean{isepjc}}%
{
\begin{figure*}
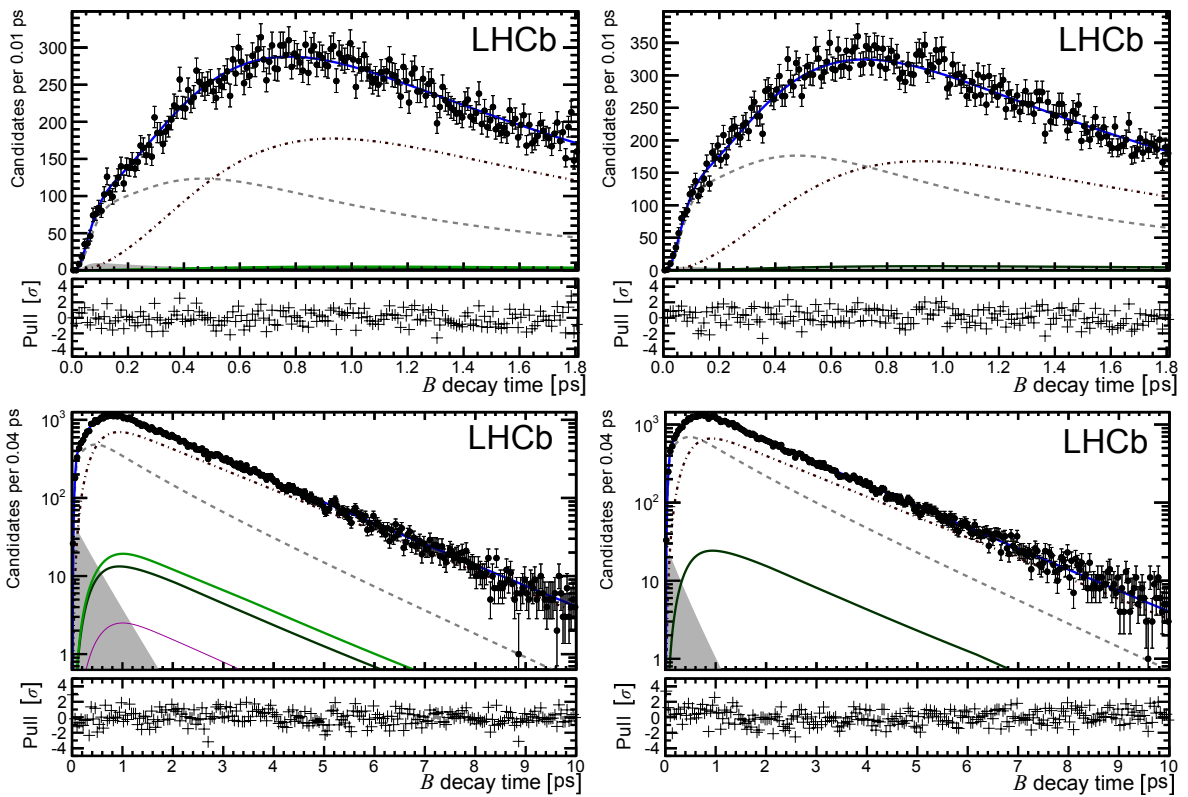

  \includegraphics[width=\columnwidth,keepaspectratio,clip=true,trim=0.4cm 0cm 0cm 1cm]{figs/Fig5a-topleft.pdf}\hspace{0.99\columnsep}
  \includegraphics[width=\columnwidth,keepaspectratio,clip=true,trim=0.4cm 0cm 0cm 1cm]{figs/Fig5b-topright.pdf}\\ \vspace{1mm}
  \includegraphics[width=0.99\columnwidth,keepaspectratio,clip=true,trim=0.4cm 0cm 0cm 1cm]{figs/Fig5c-bottomleft.pdf}\hspace{0.99\columnsep} 
  \includegraphics[width=0.99\columnwidth,keepaspectratio,clip=true,trim=0.4cm 0cm 0cm 1cm]{figs/Fig5d-bottomright.pdf} 
  \caption{Measured $B$ decay-time distribution, overlaid with projections of the fit, for background-only regions. Top left: a region between the two signal peaks, $-80$~to~$-20$\,MeV$c^{-2}$ (with respect to the known mass of the $D_s^+$), showing only low decay times. Top right: a region to the right of the signal peaks $20$~to~$100$\,MeV$c^{-2}$, showing only low decay times. Bottom row: the same on an extended decay-time scale and logarithmic. The legend is the same as in \mfig~\ref{Figure:model:mass}.}
  \label{Figure:model:tauB}
\end{figure*}
}{}

\ifthenelse{\boolean{isepjc}}%
{
\begin{figure*}
  \includegraphics[width=\columnwidth,keepaspectratio,clip=true,trim=0.4cm 0cm 0cm 1cm]{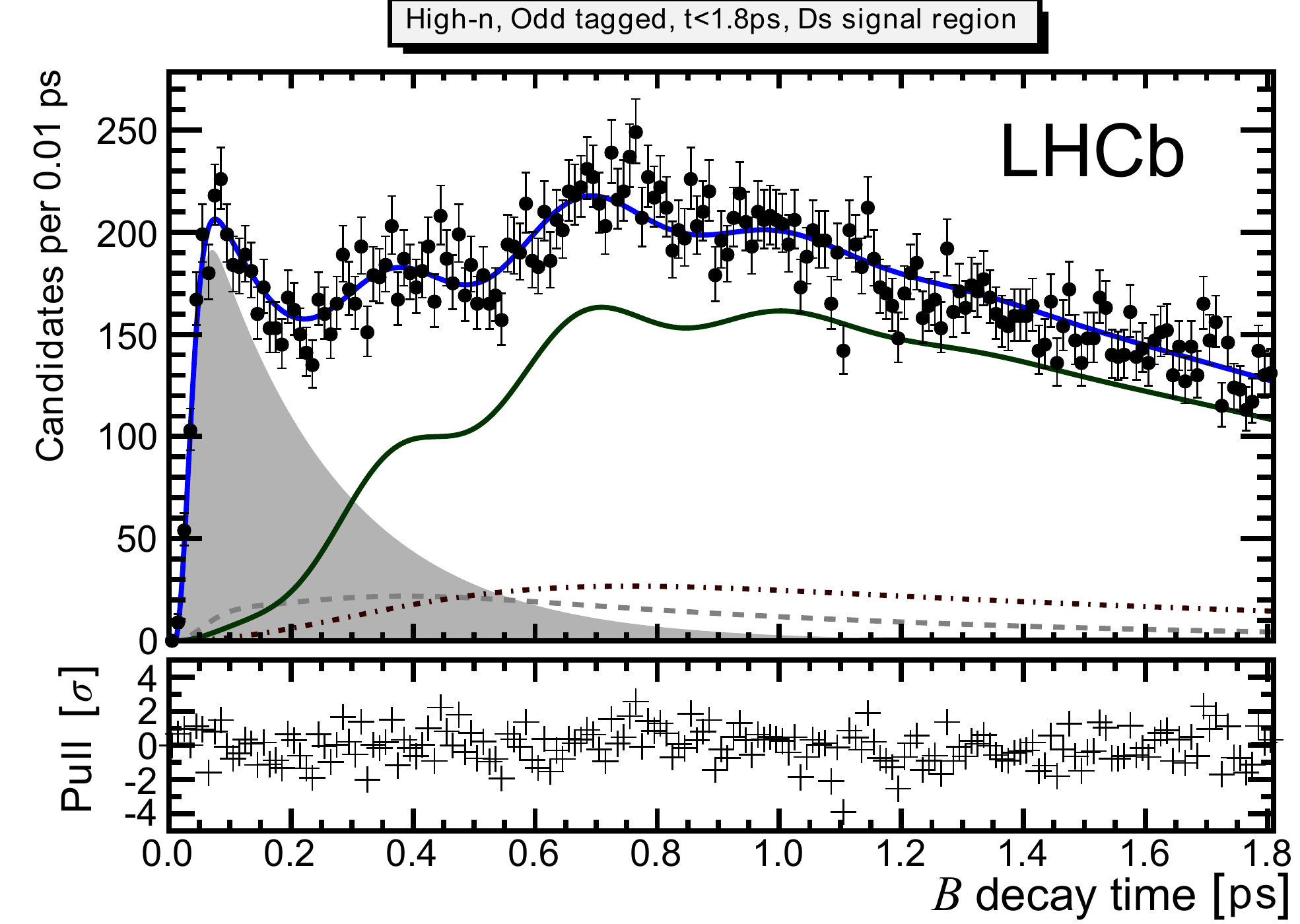}\hspace{0.99\columnsep}
  \includegraphics[width=\columnwidth,keepaspectratio,clip=true,trim=0.4cm 0cm 0cm 1cm]{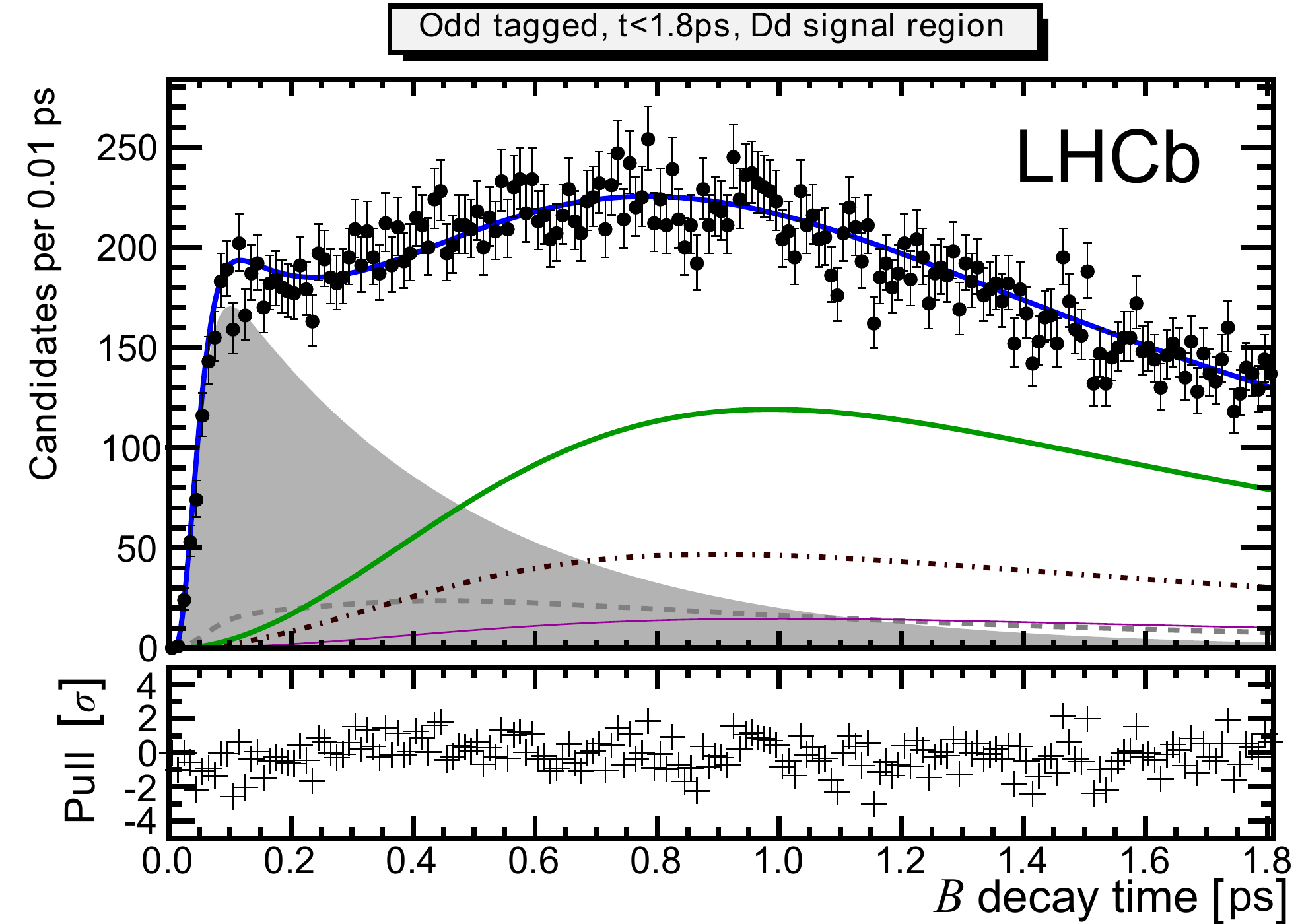}\\ \vspace{1mm}
  \includegraphics[width=0.99\columnwidth,keepaspectratio,clip=true,trim=0.4cm 0cm 0cm 1cm]{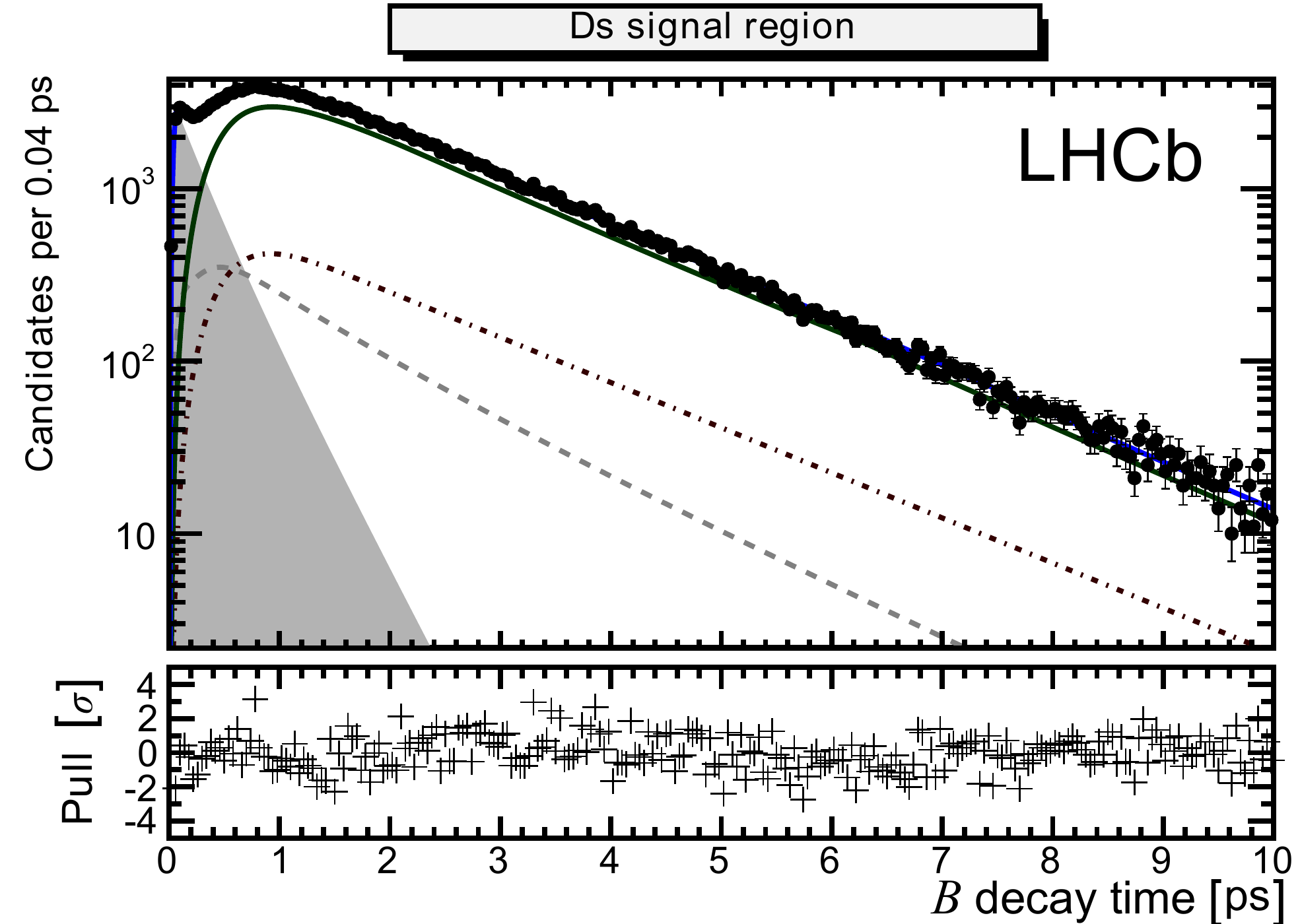}\hspace{0.99\columnsep} 
  \includegraphics[width=0.99\columnwidth,keepaspectratio,clip=true,trim=0.4cm 0cm 0cm 1cm]{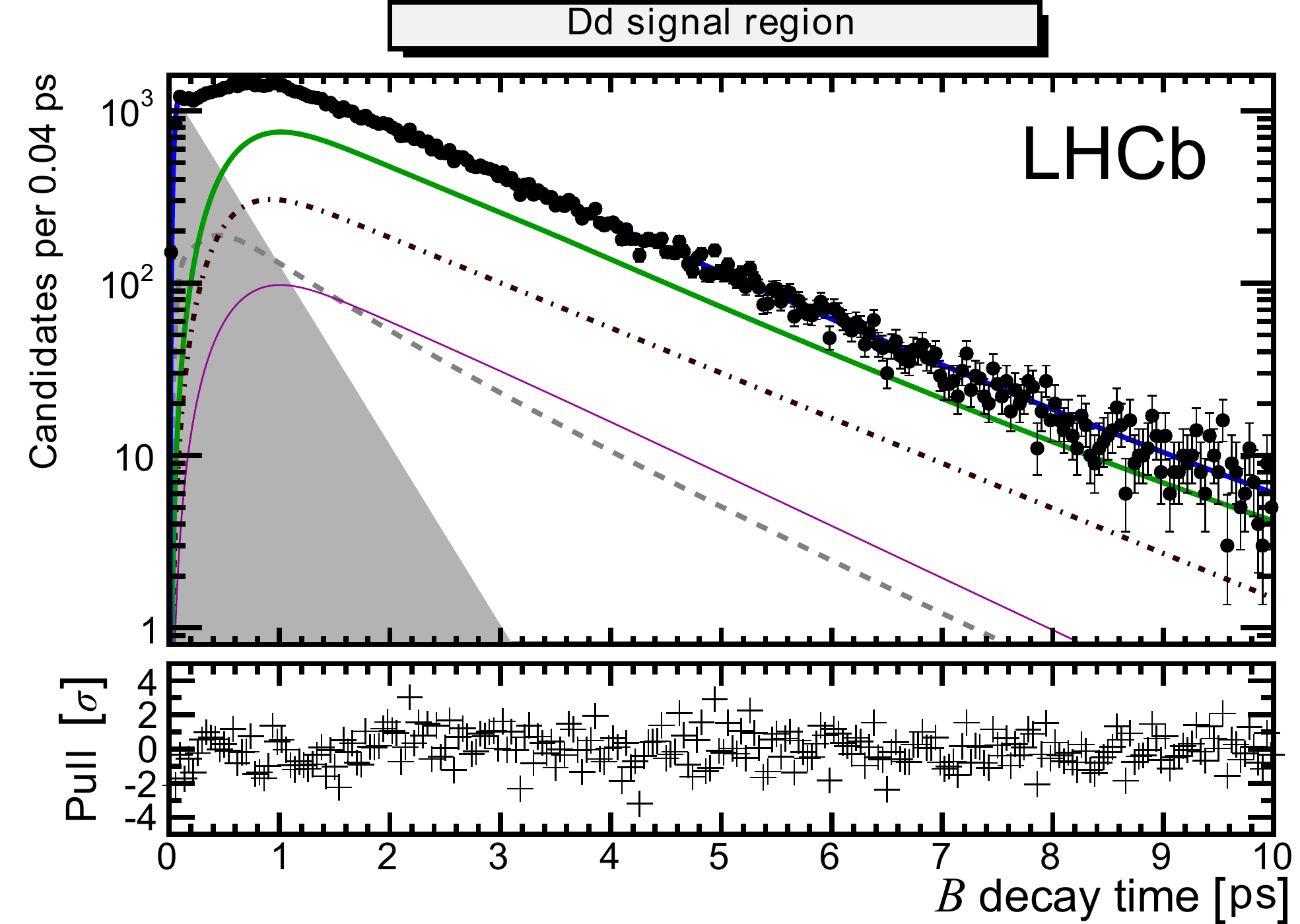} 
  \caption{Measured $B$ decay-time distribution, overlaid with projections of the fit, for signal regions. Top left: for odd-tags, high-$n$ and a region of $\pm 20$\,MeV$c^{-2}$ around the $D^+_s$ mass peak, showing only low decay times, where $B^0_s$ oscillations can be clearly seen. Top right: for odd-tags and all $n$ for a region of $\pm 20$\,MeV$c^{-2}$ around the $D^+$ mass peak, showing only low decay times. Bottom row: for both tags and all $n$ for regions of $\pm 20$\,MeV$c^{-2}$ around the $D^+_s$ (left) and $D^+$ (right) mass peaks. The legend is the same as in \mfig~\ref{Figure:model:mass}.}
  \label{Figure:model:tau}
\end{figure*}
}{
\begin{figure}[b]
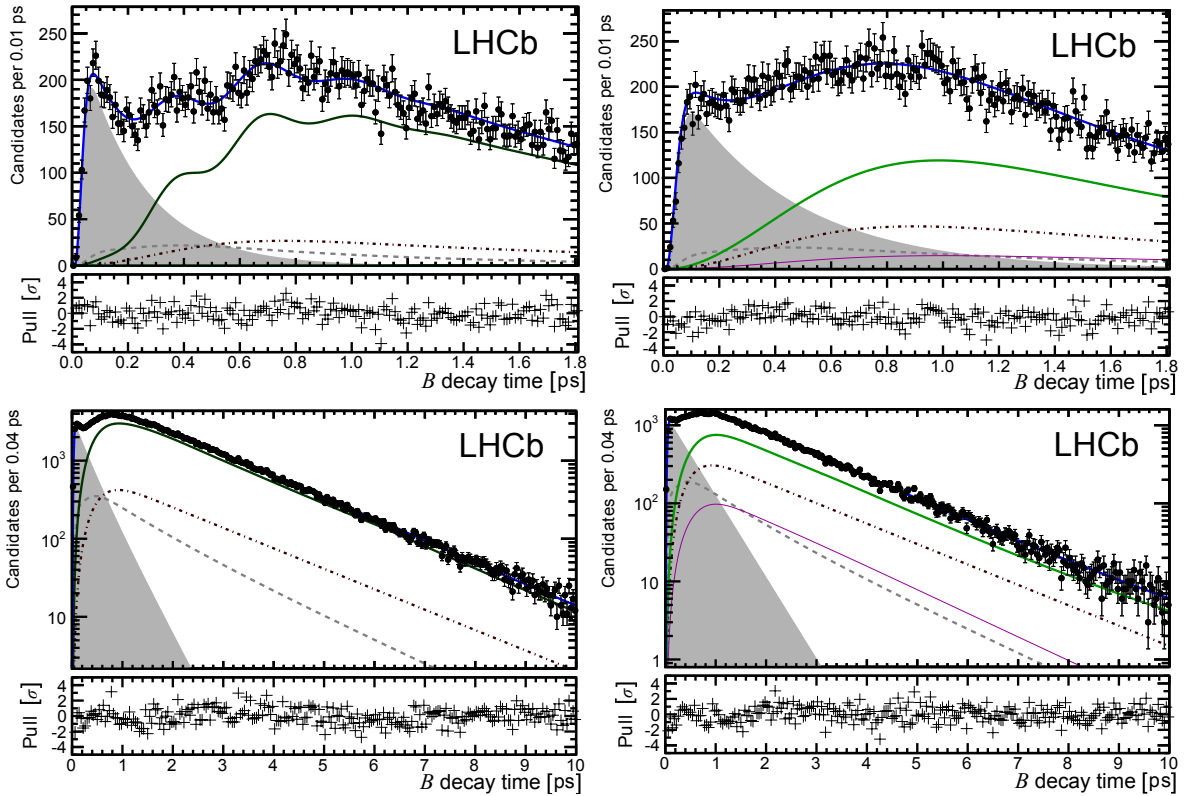

  \centering
  \includegraphics[width=0.48\textwidth,keepaspectratio,clip=true,trim=0.4cm 0cm 0cm 1cm]{figs/Fig6a-topleft.pdf}\hspace{1mm}
  \includegraphics[width=0.48\textwidth,keepaspectratio,clip=true,trim=0.4cm 0cm 0cm 1cm]{figs/Fig6b-topright.pdf}\\ \vspace{1mm}
  \includegraphics[width=0.478\textwidth,keepaspectratio,clip=true,trim=0.4cm 0cm 0cm 1cm]{figs/Fig6c-bottomleft.pdf}\hspace{0.2mm} 
  \includegraphics[width=0.478\textwidth,keepaspectratio,clip=true,trim=0.4cm 0cm 0cm 1cm]{figs/Fig6d-bottomright.pdf} 
  \caption{Measured $B$ decay-time distribution, overlaid with projections of the fit, for signal regions. Top left: for odd-tags, high-$n$ and a region of $\pm 20$\,MeV$c^{-2}$ around the $D^+_s$ mass peak, showing only low decay times, where $B^0_s$ oscillations can be clearly seen. Top right: for odd-tags and all $n$ for a region of $\pm 20$\,MeV$c^{-2}$ around the $D^+$ mass peak, showing only low decay times. Bottom row: for both tags and all $n$ for regions of $\pm 20$\,MeV$c^{-2}$ around the $D^+_s$ (left) and
$D^+$ (right) mass peaks. The legend is the same as in \mfig~\ref{Figure:model:mass}.}
  \label{Figure:model:tau}
\end{figure}
}

\ifthenelse{\boolean{isepjc}}%
{
\begin{figure*}
  \includegraphics[width=\columnwidth,keepaspectratio,clip=true,trim=0.25cm 0cm 0cm 0.9cm]{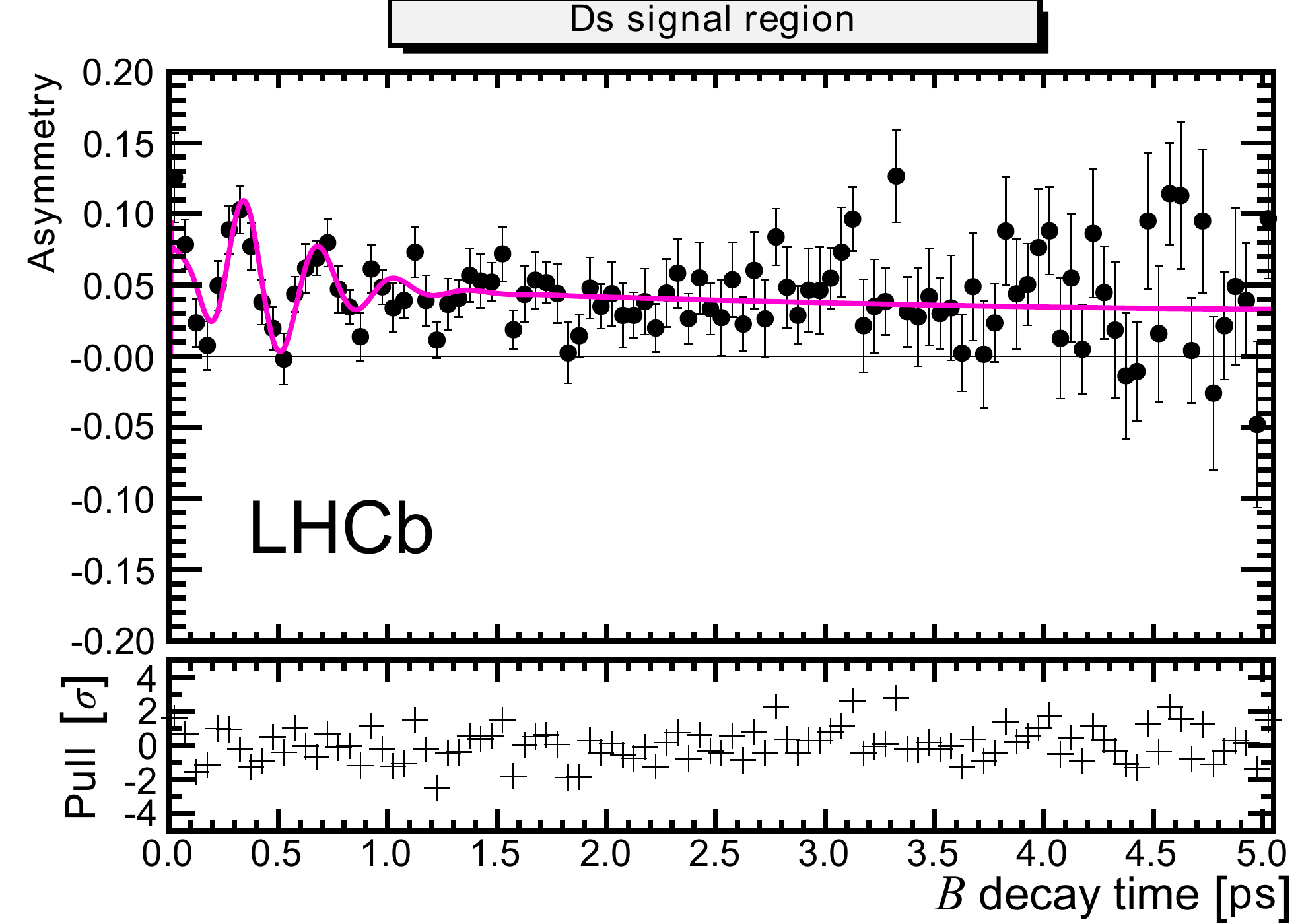}\hspace{0.99\columnsep}
  \includegraphics[width=\columnwidth,keepaspectratio,clip=true,trim=0.25cm 0cm 0cm 0.9cm]{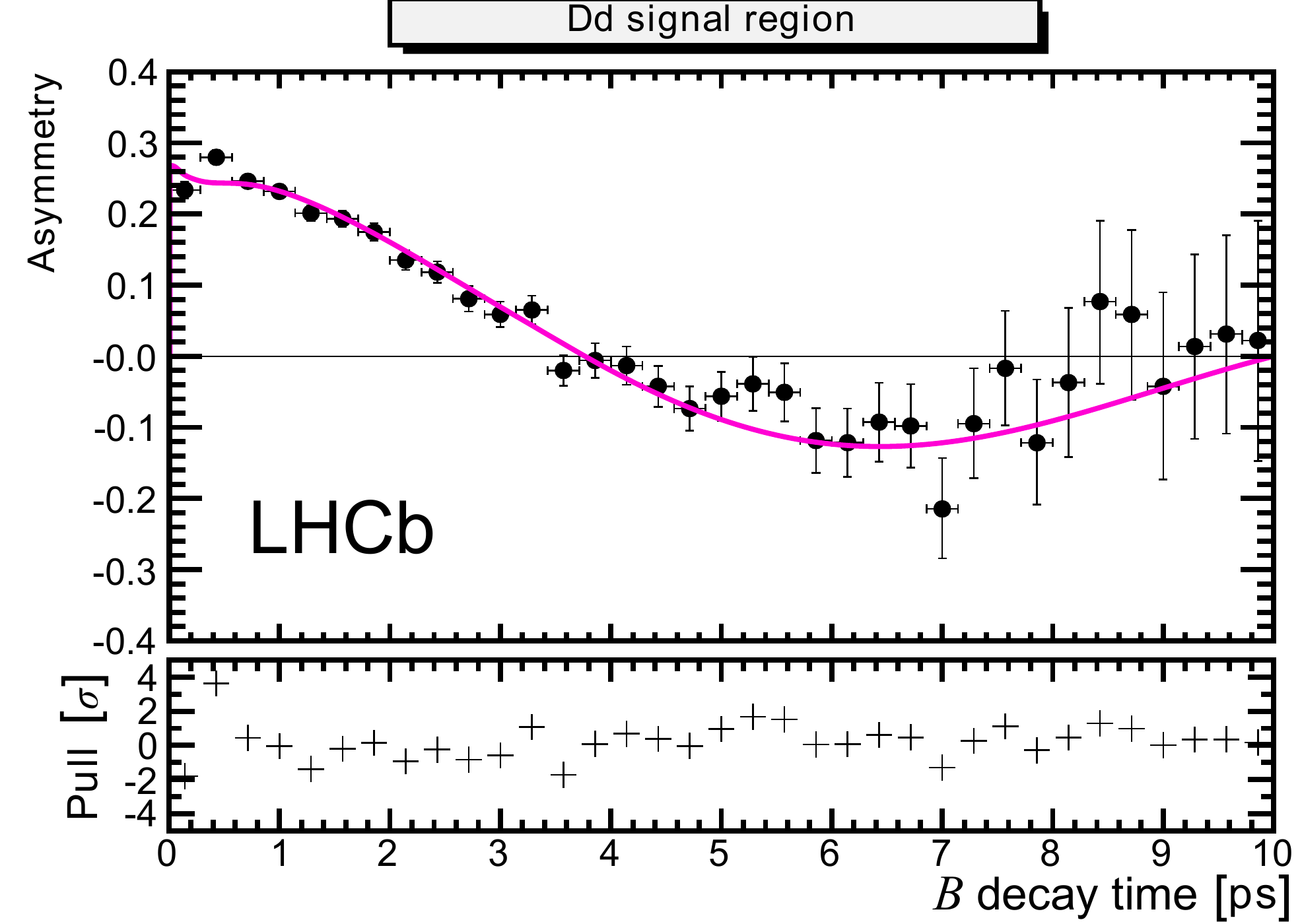}
  \caption{Tagged (mixing) asymmetry, $(N_+-N_-)/(N_++N_-)$, as a function of $B$ decay time. The left plot shows the asymmetry for events for a region of $\pm 20$\,MeV$c^{-2}$ around the $D^+_s$ mass peak, and the right plot shows the corresponding asymmetry around the $D^+$ mass peak. The black points show the data and the curves are projections of the fitted PDF. On the left plot the fast oscillations of $B^0_s$ are gradually washed out by the increasingly poor decay-time resolution.}
  \label{Figure:signal-asym}
\end{figure*}
}
{
\begin{figure}[b]
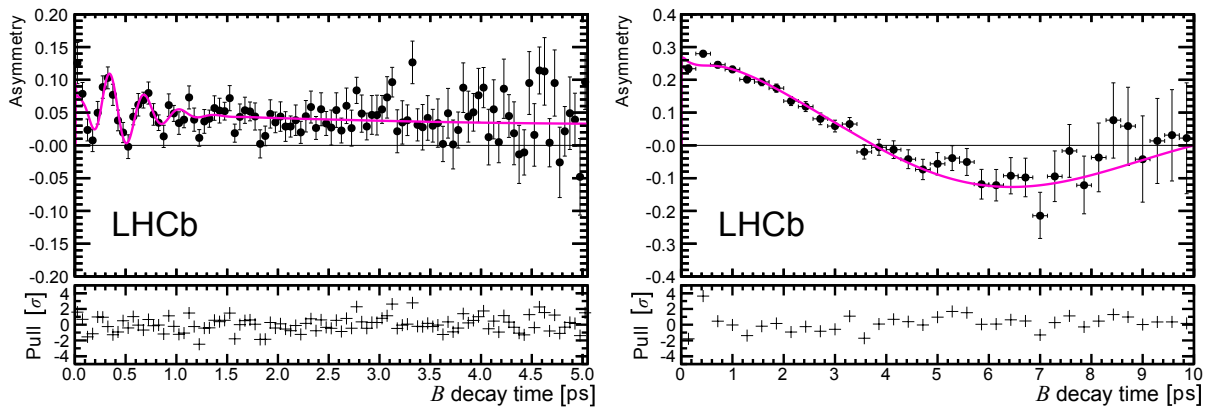

  \centering
  \includegraphics[width=0.49\textwidth,keepaspectratio,clip=true,trim=0.25cm 0cm 0cm 0.9cm]{figs/Fig7a-left.pdf}
  \includegraphics[width=0.49\textwidth,keepaspectratio,clip=true,trim=0.25cm 0cm 0cm 0.9cm]{figs/Fig7b-right.pdf}
  \caption{Tagged (mixing) asymmetry, $(N_+-N_-)/(N_++N_-)$, as a function of $B$ decay time. The left plot shows the asymmetry for events for a region of $\pm 20$\,MeV$c^{-2}$ around the $D^+_s$ mass peak, and the right plot shows the corresponding asymmetry around the $D^+$ mass peak. The black points show the data and the curves are projections of the fitted PDF. On the left plot the fast oscillations of $B^0_s$ are gradually washed out by the increasingly poor decay-time resolution.}
  \label{Figure:signal-asym}
\end{figure}
}

The proportion of $B^+ \to D^{-} \mu^{+} (\nu, \pi^+, \gamma)$ with respect to $B^0 \to D^{-} \mu^{+} (\nu, \pi^0, \gamma)$ is fixed to $11\,\%$ with a ${\pm}2\,\%$ uncertainty, using the ratio of known fragmentation functions and branching fractions~\cite{PDG2012}.  Based on the full LHCb simulation, this ratio is corrected by $25\,\%$ to account for differences in the reconstruction and tagging efficiencies, with the full value of this correction taken as
a systematic uncertainty. We fix $\Delta \Gamma_s$ using the result of a recent LHCb analysis~\cite{LHCb-PAPER-2013-002}, and $\Delta \Gamma_d$ is fixed to zero.

Only the signal mass shapes and the parameters of interest, $\Delta m_s$ and $\Delta m_d$, are shared between the two subsamples in $n$, which are fitted simultaneously. The goodness of the fit is verified with a local density method~\cite{Mike:GOF}, which finds a $p$-value of $19.6\,\%$.

\section{Fit results and systematic uncertainties}
\label{Section:Systematics}

\ifthenelse{\boolean{isepjc}}%
{ \ifthenelse{\boolean{isepjc}}%
{
\begin{table}
  \caption{A selection of fitted parameter values, for which statistical uncertainties only are given. The $B^0_s$ signal fraction includes contributions from any detached $D^+_s$ production. When the omitted fractions (of combinatorial background components) are included, the total fraction sums to unity within each $n$ region separately.}
  \begin{center}
  \label{Table:Results}
  \begin{tabular}{l|cc}\hline\noalign{\smallskip}
    Quantity &  \multicolumn{2}{c}{Normalized mass region}  \\
        & Low-$n$ & High-$n$  \\ \noalign{\smallskip}\hline\noalign{\smallskip}
      Fit fraction of:   &  &   \\
      ~-~$B^0_s$ signal  &  0.3247$\pm$0.0029 & 0.3604$\pm$0.0023   \\
      ~-~$B^0$ signal  &  0.0781$\pm$0.0017 & 0.0968$\pm$0.0022   \\
      ~-~prompt $D^+_s$  &  0.0410$\pm$0.0026 & 0.0444$\pm$0.0018   \\
      ~-~prompt $D^+$  &  0.0196$\pm$0.0018 & 0.0311$\pm$0.0024   \\
      Mistag probability $\omega$: & &\\
      ~-~$B^0_s$ signal  &  0.347$\pm$0.054 & 0.333$\pm$0.021   \\
      ~-~$B^0$ signal  &  0.3567$\pm$0.0063 & 0.3319$\pm$0.0065   \\\noalign{\smallskip}\hline\noalign{\smallskip}
      Total candidates & 368,965 & 225,880  \\ \noalign{\smallskip}\hline
      \end{tabular}
  \end{center}
\end{table}
}
{
\begin{table}[b]
  \caption{A selection of fitted parameter values, for which statistical uncertainties only are given. The $B^0_s$ signal fraction includes contributions from any detached $D^+_s$ production. When the omitted fractions (of combinatorial background components) are included, the total fraction sums to unity within each $n$ region separately.} \small
  \begin{center}
  \label{Table:Results}
  \begin{tabular}{l|cc}\hline
    Quantity &  \multicolumn{2}{c}{Normalized mass region}  \\
        & Low-$n$ & High-$n$  \\ \hline
      Fit fraction of:   &  &   \\
      ~-~$B^0_s$ signal  &  0.3247$\pm$0.0029 & 0.3604$\pm$0.0023   \\
      ~-~$B^0$ signal  &  0.0781$\pm$0.0017 & 0.0968$\pm$0.0022   \\
      ~-~prompt $D^+_s$  &  0.0410$\pm$0.0026 & 0.0444$\pm$0.0018   \\
      ~-~prompt $D^+$  &  0.0196$\pm$0.0018 & 0.0311$\pm$0.0024   \\
      Mistag probability $\omega$: & &\\
      ~-~$B^0_s$ signal  &  0.347$\pm$0.054 & 0.333$\pm$0.021   \\
      ~-~$B^0$ signal  &  0.3567$\pm$0.0063 & 0.3319$\pm$0.0065   \\\hline
      Total candidates & 368,965 & 225,880  \\ \hline
      \end{tabular}
  \end{center}
\end{table}
} }
{}

Table~\ref{Table:Results} gives the fitted values for some important quantities. In principle the signal lifetimes are also measured, but these have very large systematic uncertainties and so no results are quoted. The systematic uncertainties on $\Delta m_s$ and $\Delta m_d$ are first discussed before the final results are given.

Several sources of systematic uncertainty on the main measured quantities, $\Delta m_s$ and $\Delta m_d$, are considered, as summarized in Table~\ref{Table:Systematics}. \ifthenelse{\boolean{isepjc}}{\ifthenelse{\boolean{isepjc}}%
{
\begin{table*}
  \caption{Sources of systematic uncertainty on $\Delta m_s$ and $\Delta m_d$. ``Simulation'' implies a combination of full LHCb simulation and pseudo-experiment studies.}
  \begin{center}
  \label{Table:Systematics}
  \begin{tabular}{lc|ll}\hline\noalign{\smallskip}
    Source of uncertainty & Method & \multicolumn{2}{c}{Systematic uncertainty}  \\
       & & $\Delta m_s$ [ps$^{-1}$] & $\Delta m_d$ [ps$^{-1}$]  \\ \noalign{\smallskip}\hline\noalign{\smallskip}
      $k$-factor & Simulation  & ~0.06  & ~0.0052\\
      Detector alignment & Calibration  & ~0.03 & ~0.0008\\
      Values of $\Delta \Gamma$ & Data refit &   ~~n/a & ~0.0004\\
      Model bias & Simulation  & ~0.09& ~0.0055 \\
      Signal proper-time model  & Data refit & ~0.09& ~0.007  \\
      Other models and binning & Data refit  & ~0.05 & ~0.001\\
      $B^+$ (\BF, efficiency, tagging) & Data refit & ~~n/a  & ~0.008 \\\noalign{\smallskip}\hline\noalign{\smallskip}
      Total & Sum in quadrature & ~0.15 & ~0.013  \\
      \noalign{\smallskip}\hline
      \end{tabular}
  \end{center}
\end{table*}
}
{
\begin{table}[b]
  \caption{Sources of systematic uncertainty on $\Delta m_s$ and $\Delta m_d$. ``Simulation'' implies a combination of full LHCb simulation and pseudo-experiment studies.} \small
  \begin{center}
  \label{Table:Systematics}
  \begin{tabular}{lc|ll}\hline
    Source of uncertainty & Method & \multicolumn{2}{c}{Systematic uncertainty}  \\
       & & $\Delta m_s$ [ps$^{-1}$] & $\Delta m_d$ [ps$^{-1}$]  \\ \hline
      $k$-factor & Simulation  & ~0.06  & ~0.0052\\
      Detector alignment & Calibration  & ~0.03 & ~0.0008\\
      Values of $\Delta \Gamma$ & Data refit &   ~~n/a & ~0.0004\\
      Model bias & Simulation  & ~0.09& ~0.0055 \\
      Signal proper-time model  & Data refit & ~0.09& ~0.007  \\
      Other models and binning & Data refit  & ~0.05 & ~0.001\\
      $B^+$ (\BF, efficiency, tagging) & Data refit & ~~n/a  & ~0.008 \\\hline
      Total & Sum in quadrature & ~0.15 & ~0.013  \\
      \hline
      \end{tabular}
  \end{center}
\end{table}
}}{}The majority of the systematic uncertainties are obtained from the data.\begin{itemize}
  \item The $k$-factor: the $k$-factor correction is a simulation-based method, and so differences between the simulation and reality that modify the visible and invisible momenta potentially invalidate the correction. Such differences could for example be in $D^{**}$ branching fractions or form factors. Large-scale pseudo-experiment studies are combined with full simulations to vary these underlying distributions within their uncertainties and examine biases produced on the fitted $\Delta m$ values. Small relative uncertainties are found, $0.3\,\%$ for $\Delta m_s$ and $1.0$\,\% for $\Delta m_d$, representing the ultimate limit of this technique without further knowledge of the various sub-decays.
  \item Detector alignment: momentum scale, decay-length scale, and track position uncertainties arise from known alignment uncertainties and result in variations in reconstructed masses and lifetimes as functions of decay opening angle. These uncertainties have been studied using detector survey data and various control modes; they are well determined and small in comparison to the statistical uncertainties~\cite{Amoraal:2012qn}.
  \item Values of $\Delta \Gamma$: The quantities $\Delta \Gamma_d$ and $\Delta \Gamma_s$ are nominally constant in our fits. When they are varied, within $\pm 5$\,\% for $\Delta \Gamma_d$ (chosen to well-cover the experimental range given the lack of information on its sign \cite{PDG2012}) and within the known uncertainty on $\Delta \Gamma_s$~\cite{LHCb-PAPER-2013-002}, our result is only marginally affected.
  \item Model bias: a correction has been made for the 1\,\% residual frequency bias seen in full simulation studies, as discussed in \sect~\ref{Section:Time}. This is taken directly from simulation and half of the correction is assigned a systematic uncertainty.
  \item Signal proper-time model: the fit is repeated with two different time-resolution models. (a) When the resolution is parameterized as a function of true rather than measured decay time, using full numerical convolution, a (0.09, 0.002)\,ps$^{-1}$ variation is seen in ($\Delta m_s$, $\Delta m_d$). (b) When a time-independent (average) resolution is used, a 0.001\,ps$^{-1}$ variation is seen in $\Delta m_d$ (this method is not applicable to the measurement of $\Delta m_s$ due to many factors; crucially, within the time frame of any single $B^0_s$ oscillation the decay time resolution worsens by an appreciable fraction of the oscillation period, seen in \mfigs~\ref{Figure:Toys:ResolutionBd} and \ref{Figure:signal-asym}). With other modifications to the signal model (resolutions and acceptances) a larger variation in $\Delta m_d$ of $0.007$\,ps$^{-1}$ is found. \ifthenelse{\boolean{isepjc}}{}{}
  \item Other models and binning: the order of the Chebychev polynomial is varied, Crystal Ball functions are used for the mass peak shapes, and the background parameterizations and the binning schemes are varied. Out of these modifications, the binning scheme has the largest effect. Resulting uncertanties of $0.05$\,ps$^{-1}$ and $0.001$\,ps$^{-1}$ are assigned to $\Delta m_s$ and $\Delta m_d$, respectively.
  \item Assumptions on $B^+$ decays: The $\Delta m_d$ measurement is sensitive to $\chi_d$, the integrated mixing probability, which in turn is sensitive to the non-mixing $B^+$-background. We hold constant several $B^+$-background parameters in the baseline fit, determined from the full simulation. Many features of the $B^+$ background fit are varied to evaluate systematic variations, including the fraction, the lifetime, and the corrections for relative tagging performance. The largest uncertainty arises from tagging performance corrections and for this a $0.008$\,ps$^{-1}$ uncertainty is assigned to $\Delta m_d$. It is possible to leave one or more of these parameters free during the fit, but the loss in statistical precision is prohibitive.
\end{itemize}

\ifthenelse{\boolean{isepjc}}{}{} For cross-checks the data are split by LHCb magnet polarity and LHCb trigger strategies; no variations beyond the expected statistical fluctuations are observed. We obtain
\begin{align}
&\Delta m_s = ( 17.93  \pm  0.22\,\textrm{(stat)} \pm  0.15\,\textrm{(syst)}) \,\textrm{ps}^{-1} \nonumber ,\\
&\Delta m_d = ( 0.503  \pm  0.011\,\textrm{(stat)} \pm  0.013\,\textrm{(syst)}) \,\textrm{ps}^{-1}\nonumber .
\end{align}

To obtain a measure for the significance of the observed oscillations, the global likelihood minimum for the full fit is compared with the likelihood of the hypotheses corresponding to the edges of our search window ($\Delta m =0$ or $\Delta m \geq 50$\,ps$^{-1}$). Both would result in almost flat asymmetry curves (cf.~\mfig~\ref{Figure:signal-asym}) corresponding to no observed oscillations. We reject the null hypothesis of no oscillations by the equivalent of $5.8$ standard deviations for $B^0_s$ oscillations and $13.0$ standard deviations for $B^0$ oscillations.


\section{Fourier analysis}
\label{Section:Fourier}

\ifthenelse{\boolean{isepjc}}%
{
\begin{figure*}
\begin{center}
\includegraphics[height=0.95\columnwidth,keepaspectratio,clip=true,trim=0cm 0cm 0cm 1cm]{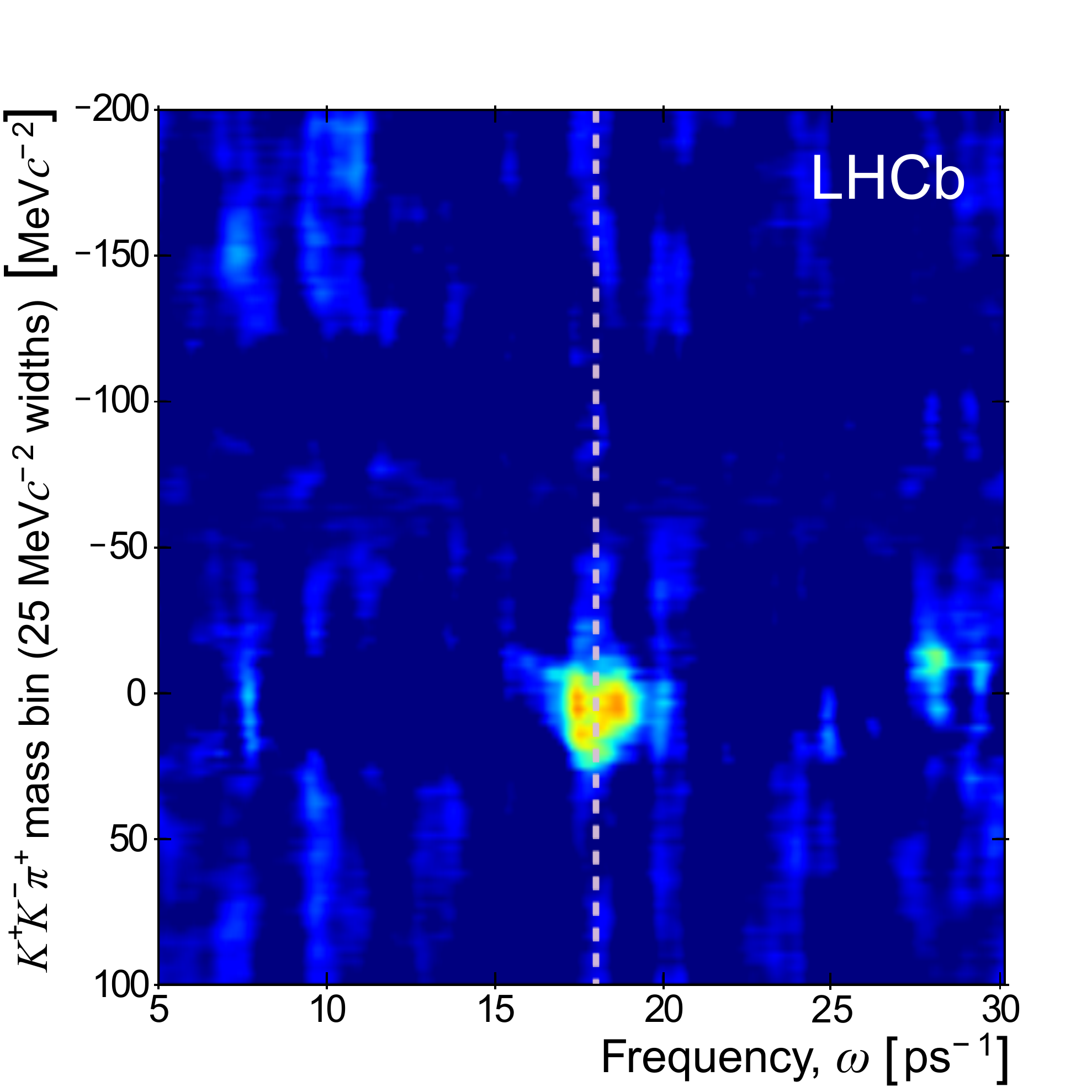}\hspace{0.99\columnsep}
  \includegraphics[height=0.95\columnwidth,keepaspectratio,clip=true,trim=24cm 0.05cm 4cm 1.35cm]{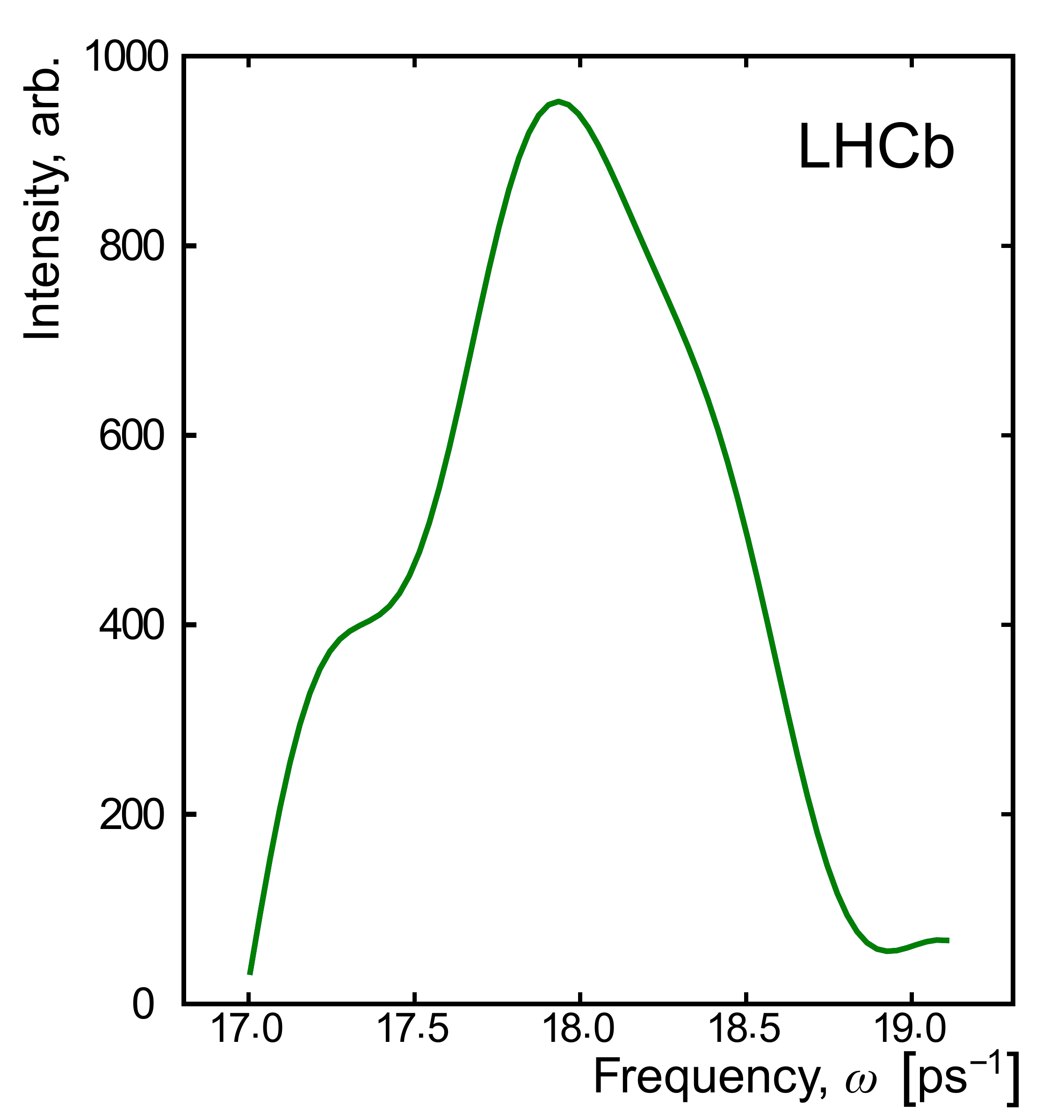}
\end{center}
 \caption{Result of using Fourier transforms to search for the $\Delta m_s$-peak. The image on the left is constructed from bins of the $K^+K^-\pi^+$ mass which are 25\,MeV$c^{-2}$ in width, analysed in steps of 5\,MeV$c^{-2}$ such that a smooth image is produced. The colour scale (blue-green-yellow-red) is an arbitrary linear representation of the signal intensity; dark blue is used for zero and below. The vertical dashed line is drawn at $18.0$\,ps$^{-1}$. The apparent double-peak structure is an artifact of this image. On the right a slice around the $D^+_s$ mass region shows only the peak as used to measure the central value and rms width.}
  \label{Figure:Fourier:Dms}
\end{figure*}
}
{ 
\begin{figure}[b]
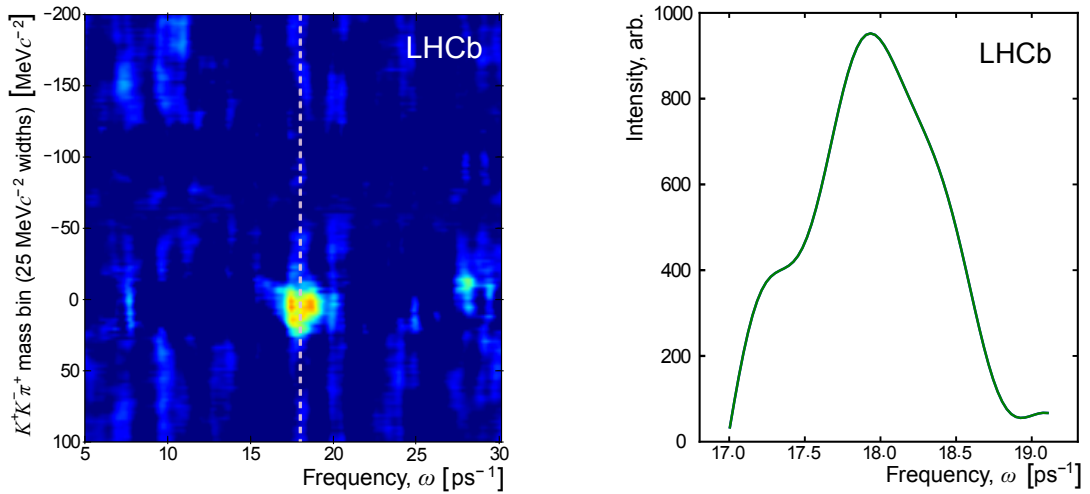

\begin{center}
\includegraphics[height=0.42\textwidth,keepaspectratio,clip=true,trim=0cm 0cm 0cm 1cm]{figs/Fig8a-left.pdf}\hspace{0.05\textwidth}
  \includegraphics[height=0.42\textwidth,keepaspectratio,clip=true,trim=24cm 0.05cm 4cm 1.35cm]{figs/Fig8b-right.pdf}
\end{center}
 \caption{Result of using Fourier transforms to search for the $\Delta m_s$-peak. The image on the left is constructed from bins of the $K^+K^-\pi^+$ mass which are 25\,MeV$c^{-2}$ in width, analysed in steps of 5\,MeV$c^{-2}$ such that a smooth image is produced. The colour scale (blue-green-yellow-red) is an arbitrary linear representation of the signal intensity; dark blue is used for zero and below. The vertical dashed line is drawn at $18.0$\,ps$^{-1}$. The apparent double-peak structure is an artifact of this image. On the right a slice around the $D^+_s$ mass region shows only the peak as used to measure the central value and rms width.}
  \label{Figure:Fourier:Dms}
\end{figure}
}

The full fit as described above was performed in the time domain, but measurement of the mixing frequency can also be made directly in the frequency domain as a cross-check, using well-established Fourier transform techniques~\cite{Fourier,FourierAnalysis,Moser:1996xf}. The cosine term in Eq.~\ref{Eqn:Intro:TaggedRes} has a different sign for the odd and even samples, where the lifetime, acceptance, and other features are shared; this simplifies the analysis in the frequency domain. Any Fourier components not arising from mixing are suppressed by subtracting the odd Fourier spectrum from the even spectrum and no parameterizations of the background shapes, signal shapes, or decay-time resolution are required, allowing a model-independent measurement of the mixing frequencies. We search for the $\Delta m_s$ peak in the subtracted Fourier spectrum, shown in \mfig~\ref{Figure:Fourier:Dms}. Extensive fast simulation pseudo-experiments have shown that the value of $\Delta m_s$ is obtained reliably and with a reasonable precision using this method; however $\Delta m_d$ is heavily biased and has a large uncertainty, and so a result is not quoted. Since residual components of the Fourier spectrum are of much lower frequency than the $\Delta m_s$ component, and several complete oscillation periods of $\Delta m_s$ are observable, the search for a spectral peak is relatively free from complications. For $\Delta m_d$, however, the relatively low frequency is similar to that of many other features of the data, and only a single oscillation period is observed; therefore the determination of $\Delta m_d$ is difficult with this simple model-independent approach.

Taking the spectrum for events in a 25\,MeV$c^{-2}$ bin around the $D^+_s$ mass, we find a clear and separated peak (\mfig~\ref{Figure:Fourier:Dms}, right). The rms~width of the peak is 0.4\,ps$^{-1}$, around a peak value of $17.95$\,ps$^{-1}$; the rms~can be used as a model-independent proxy for the statistical uncertainty. To further evaluate the expected statistical fluctuation in the peak value, we perform a large set of fast simulation pseudo-experiments taking the result of the multivariate fit as a model for signal and background. The uncertainty found from the simulation studies is 0.32\,ps$^{-1}$, slightly smaller than given by the rms. We report $\Delta m_s =(17.95 \pm 0.40\,\textrm{(rms)} \pm 0.11\,\textrm{(syst)})$\,ps$^{-1},$ in order to be model-independent. Systematic uncertainties arise from the detector alignment and the $k$-factor correction method, common to both measurement techniques, as quantified previously in \sect~\ref{Section:Systematics}.


\section{Conclusion}

The mixing frequencies for neutral $B$ mesons have been measured using flavour-specific semileptonic decays. To correct for the momentum lost to missing particles, a simulation-based kinematic correction, known as the $k$-factor, was adopted. Two techniques were used to measure the mixing frequencies:  a multidimensional simultaneous fit to the $K^+K^-\pi^+$ mass distribution, the decay-time distribution, and tagging information; and a simple Fourier analysis. The results of the two methods were consistent, with the first method being more precise. We obtain
\begin{align}
&\Delta m_s = ( 17.93  \pm  0.22\,\textrm{(stat)} \pm  0.15\,\textrm{(syst)}) \,\textrm{ps}^{-1} \nonumber ,\\
&\Delta m_d = ( 0.503  \pm  0.011\,\textrm{(stat)} \pm  0.013\,\textrm{(syst)}) \,\textrm{ps}^{-1}\nonumber .
\end{align}

We reject the hypothesis of no oscillations by 5.8 standard deviations for $B^0_s$ and 13.0 standard deviations for $B^0$. This is the first observation of $B^0_s$-\Bsb mixing to be made using only semileptonic decays.

\ifthenelse{\boolean{isepjc}}%
{
~\\\noindent \small \textbf{Acknowledgements} ~~
}{\section*{Acknowledgements} \noindent}We express our gratitude to our colleagues in the CERN
accelerator departments for the excellent performance of the LHC. We
thank the technical and administrative staff at the LHCb
institutes. We acknowledge support from CERN and from the national
agencies: CAPES, CNPq, FAPERJ and FINEP (Brazil); NSFC (China);
CNRS/IN2P3 and Region Auvergne (France); BMBF, DFG, HGF and MPG
(Germany); SFI (Ireland); INFN (Italy); FOM and NWO (The Netherlands);
SCSR (Poland); MEN/IFA (Romania); MinES, Rosatom, RFBR and NRC
``Kurchatov Institute'' (Russia); MinECo, XuntaGal and GENCAT (Spain);
SNSF and SER (Switzerland); NAS Ukraine (Ukraine); STFC (United
Kingdom); NSF (USA). We also acknowledge the support received from the
ERC under FP7. The Tier1 computing centres are supported by IN2P3
(France), KIT and BMBF (Germany), INFN (Italy), NWO and SURF (The
Netherlands), PIC (Spain), GridPP (United Kingdom). We are thankful
for the computing resources put at our disposal by Yandex LLC
(Russia), as well as to the communities behind the multiple open
source software packages that we depend on.

\addcontentsline{toc}{section}{References}
\setboolean{inbibliography}{true}
\bibliographystyle{LHCb}

\begin{mcitethebibliography}{10}
\mciteSetBstSublistMode{n}
\mciteSetBstMaxWidthForm{subitem}{\alph{mcitesubitemcount})}
\mciteSetBstSublistLabelBeginEnd{\mcitemaxwidthsubitemform\space}
{\relax}{\relax}

\bibitem{PDG2012}
Particle Data Group, J.~Beringer {\em et~al.},
  \ifthenelse{\boolean{articletitles}}{{\it {\href{http://pdg.lbl.gov/}{Review
  of particle physics}}},
  }{}\href{http://dx.doi.org/10.1103/PhysRevD.86.010001}{Phys.\ Rev.\  {\bf
  D86} (2012) 010001}\relax
\mciteBstWouldAddEndPuncttrue
\mciteSetBstMidEndSepPunct{\mcitedefaultmidpunct}
{\mcitedefaultendpunct}{\mcitedefaultseppunct}\relax
\EndOfBibitem
\bibitem{Schneider:B0mixB:2006}
O.~Schneider, for the Particle Data Group, \emph{$B^{0}$ - $\bar{B}^{0}$
  mixing,} in \cite{PDG2012}.\relax
\mciteBstWouldAddEndPunctfalse
\mciteSetBstMidEndSepPunct{\mcitedefaultmidpunct}
{}{\mcitedefaultseppunct}\relax
\EndOfBibitem
\bibitem{Albrecht:1987dr}
ARGUS collaboration, H.~Albrecht {\em et~al.},
  \ifthenelse{\boolean{articletitles}}{{\it {Observation of $B^0$ - $\bar{B}^0$
  mixing}}, }{}\href{http://dx.doi.org/10.1016/0370-2693(87)91177-4}{Phys.\
  Lett.\  {\bf B192} (1987) 245}\relax
\mciteBstWouldAddEndPuncttrue
\mciteSetBstMidEndSepPunct{\mcitedefaultmidpunct}
{\mcitedefaultendpunct}{\mcitedefaultseppunct}\relax
\EndOfBibitem
\bibitem{Abulencia:2006ze}
CDF collaboration, A.~Abulencia {\em et~al.},
  \ifthenelse{\boolean{articletitles}}{{\it {Observation of $B^0_s$ -
  $\bar{B}^0_s$ oscillations}},
  }{}\href{http://dx.doi.org/10.1103/PhysRevLett.97.242003}{Phys.\ Rev.\ Lett.\
   {\bf 97} (2006) 242003}, \href{http://arxiv.org/abs/hep-ex/0609040}{{\tt
  arXiv:hep-ex/0609040}}\relax
\mciteBstWouldAddEndPuncttrue
\mciteSetBstMidEndSepPunct{\mcitedefaultmidpunct}
{\mcitedefaultendpunct}{\mcitedefaultseppunct}\relax
\EndOfBibitem
\bibitem{LHCb-PAPER-2013-006}
LHCb collaboration, R.~Aaij {\em et~al.},
  \ifthenelse{\boolean{articletitles}}{{\it {Precision measurement of the
  $B^0_s$--$\overline{B}^0_s$ oscillation frequency $\Delta m_s$ in the decay
  $B^0_s \to D^+_s \pi^-$}},
  }{}\href{http://dx.doi.org/10.1088/1367-2630/15/5/053021}{New J.\ Phys.\
  {\bf 15} (2013) 053021}, \href{http://arxiv.org/abs/1304.4741}{{\tt
  arXiv:1304.4741}}\relax
\mciteBstWouldAddEndPuncttrue
\mciteSetBstMidEndSepPunct{\mcitedefaultmidpunct}
{\mcitedefaultendpunct}{\mcitedefaultseppunct}\relax
\EndOfBibitem
\bibitem{LHCb-PAPER-2011-027}
LHCb collaboration, R.~Aaij {\em et~al.},
  \ifthenelse{\boolean{articletitles}}{{\it {Opposite-side flavour tagging of
  \B mesons at the LHCb experiment}},
  }{}\href{http://dx.doi.org/10.1140/epjc/s10052-012-2022-1}{Eur.\ Phys.\ J.\
  {\bf C72} (2012) 2022}, \href{http://arxiv.org/abs/1202.4979}{{\tt
  arXiv:1202.4979}}\relax
\mciteBstWouldAddEndPuncttrue
\mciteSetBstMidEndSepPunct{\mcitedefaultmidpunct}
{\mcitedefaultendpunct}{\mcitedefaultseppunct}\relax
\EndOfBibitem
\bibitem{LHCB-PAPER-2012-032}
LHCb collaboration, R.~Aaij {\em et~al.},
  \ifthenelse{\boolean{articletitles}}{{\it {Measurement of the
  $B^0$--$\overline{B}^0$ oscillation frequency $\Delta m_d$ with the decays
  $B^0 \to D^- \pi^+$ and $B^0 \to J/\psi K^{\ast 0}$}},
  }{}\href{http://dx.doi.org/10.1016/j.physletb.2013.01.019}{Phys.\ Lett.\
  {\bf B719} (2013) 318}, \href{http://arxiv.org/abs/1210.6750}{{\tt
  arXiv:1210.6750}}\relax
\mciteBstWouldAddEndPuncttrue
\mciteSetBstMidEndSepPunct{\mcitedefaultmidpunct}
{\mcitedefaultendpunct}{\mcitedefaultseppunct}\relax
\EndOfBibitem
\bibitem{Alves:2008zz}
LHCb collaboration, A.~A. Alves~Jr. {\em et~al.},
  \ifthenelse{\boolean{articletitles}}{{\it {The \lhcb detector at the LHC}},
  }{}\href{http://dx.doi.org/10.1088/1748-0221/3/08/S08005}{JINST {\bf 3}
  (2008) S08005}\relax
\mciteBstWouldAddEndPuncttrue
\mciteSetBstMidEndSepPunct{\mcitedefaultmidpunct}
{\mcitedefaultendpunct}{\mcitedefaultseppunct}\relax
\EndOfBibitem
\bibitem{LHCb-DP-2012-003}
M.~Adinolfi {\em et~al.}, \ifthenelse{\boolean{articletitles}}{{\it
  {Performance of the \lhcb RICH detector at the LHC}},
  }{}\href{http://dx.doi.org/10.1140/epjc/s10052-013-2431-9}{Eur.\ Phys.\ J.\
  {\bf C73} (2013) 2431}, \href{http://arxiv.org/abs/1211.6759}{{\tt
  arXiv:1211.6759}}\relax
\mciteBstWouldAddEndPuncttrue
\mciteSetBstMidEndSepPunct{\mcitedefaultmidpunct}
{\mcitedefaultendpunct}{\mcitedefaultseppunct}\relax
\EndOfBibitem
\bibitem{LHCb-DP-2012-002}
A.~A. Alves~Jr {\em et~al.}, \ifthenelse{\boolean{articletitles}}{{\it
  {Performance of the LHCb muon system}},
  }{}\href{http://dx.doi.org/10.1088/1748-0221/8/02/P02022}{JINST {\bf 8}
  (2013) P02022}, \href{http://arxiv.org/abs/1211.1346}{{\tt
  arXiv:1211.1346}}\relax
\mciteBstWouldAddEndPuncttrue
\mciteSetBstMidEndSepPunct{\mcitedefaultmidpunct}
{\mcitedefaultendpunct}{\mcitedefaultseppunct}\relax
\EndOfBibitem
\bibitem{LHCb-DP-2012-004}
R.~Aaij {\em et~al.}, \ifthenelse{\boolean{articletitles}}{{\it {The \lhcb
  trigger and its performance in 2011}},
  }{}\href{http://dx.doi.org/10.1088/1748-0221/8/04/P04022}{JINST {\bf 8}
  (2013) P04022}, \href{http://arxiv.org/abs/1211.3055}{{\tt
  arXiv:1211.3055}}\relax
\mciteBstWouldAddEndPuncttrue
\mciteSetBstMidEndSepPunct{\mcitedefaultmidpunct}
{\mcitedefaultendpunct}{\mcitedefaultseppunct}\relax
\EndOfBibitem
\bibitem{Sjostrand:2006za}
{Sj\"{o}strand, Torbj\"{o}rn and Mrenna, Stephen and Skands, Peter},
  \ifthenelse{\boolean{articletitles}}{{\it {PYTHIA 6.4 physics and manual}},
  }{}\href{http://dx.doi.org/10.1088/1126-6708/2006/05/026}{JHEP {\bf 05}
  (2006) 026}, \href{http://arxiv.org/abs/hep-ph/0603175}{{\tt
  arXiv:hep-ph/0603175}}\relax
\mciteBstWouldAddEndPuncttrue
\mciteSetBstMidEndSepPunct{\mcitedefaultmidpunct}
{\mcitedefaultendpunct}{\mcitedefaultseppunct}\relax
\EndOfBibitem
\bibitem{LHCb-PROC-2010-056}
I.~Belyaev {\em et~al.}, \ifthenelse{\boolean{articletitles}}{{\it {Handling of
  the generation of primary events in \gauss, the \lhcb simulation framework}},
  }{}\href{http://dx.doi.org/10.1109/NSSMIC.2010.5873949}{Nuclear Science
  Symposium Conference Record (NSS/MIC) {\bf IEEE} (2010) 1155}\relax
\mciteBstWouldAddEndPuncttrue
\mciteSetBstMidEndSepPunct{\mcitedefaultmidpunct}
{\mcitedefaultendpunct}{\mcitedefaultseppunct}\relax
\EndOfBibitem
\bibitem{Lange:2001uf}
D.~J. Lange, \ifthenelse{\boolean{articletitles}}{{\it {The EvtGen particle
  decay simulation package}},
  }{}\href{http://dx.doi.org/10.1016/S0168-9002(01)00089-4}{Nucl.\ Instrum.\
  Meth.\  {\bf A462} (2001) 152}\relax
\mciteBstWouldAddEndPuncttrue
\mciteSetBstMidEndSepPunct{\mcitedefaultmidpunct}
{\mcitedefaultendpunct}{\mcitedefaultseppunct}\relax
\EndOfBibitem
\bibitem{Golonka:2005pn}
P.~Golonka and Z.~Was, \ifthenelse{\boolean{articletitles}}{{\it {PHOTOS Monte
  Carlo: a precision tool for QED corrections in $Z$ and $W$ decays}},
  }{}\href{http://dx.doi.org/10.1140/epjc/s2005-02396-4}{Eur.\ Phys.\ J.\  {\bf
  C45} (2006) 97}, \href{http://arxiv.org/abs/hep-ph/0506026}{{\tt
  arXiv:hep-ph/0506026}}\relax
\mciteBstWouldAddEndPuncttrue
\mciteSetBstMidEndSepPunct{\mcitedefaultmidpunct}
{\mcitedefaultendpunct}{\mcitedefaultseppunct}\relax
\EndOfBibitem
\bibitem{Allison:2006ve}
Geant4 collaboration, J.~Allison {\em et~al.},
  \ifthenelse{\boolean{articletitles}}{{\it {Geant4 developments and
  applications}}, }{}\href{http://dx.doi.org/10.1109/TNS.2006.869826}{IEEE
  Trans.\ Nucl.\ Sci.\  {\bf 53} (2006) 270}\relax
\mciteBstWouldAddEndPuncttrue
\mciteSetBstMidEndSepPunct{\mcitedefaultmidpunct}
{\mcitedefaultendpunct}{\mcitedefaultseppunct}\relax
\EndOfBibitem
\bibitem{Agostinelli:2002hh}
Geant4 collaboration, S.~Agostinelli {\em et~al.},
  \ifthenelse{\boolean{articletitles}}{{\it {Geant4: a simulation toolkit}},
  }{}\href{http://dx.doi.org/10.1016/S0168-9002(03)01368-8}{Nucl.\ Instrum.\
  Meth.\  {\bf A506} (2003) 250}\relax
\mciteBstWouldAddEndPuncttrue
\mciteSetBstMidEndSepPunct{\mcitedefaultmidpunct}
{\mcitedefaultendpunct}{\mcitedefaultseppunct}\relax
\EndOfBibitem
\bibitem{LHCb-PROC-2011-006}
M.~Clemencic {\em et~al.}, \ifthenelse{\boolean{articletitles}}{{\it {The \lhcb
  simulation application, \gauss: design, evolution and experience}},
  }{}\href{http://dx.doi.org/10.1088/1742-6596/331/3/032023}{{J.\ Phys.\ \!\!:
  Conf.\ Ser.\ } {\bf 331} (2011) 032023}\relax
\mciteBstWouldAddEndPuncttrue
\mciteSetBstMidEndSepPunct{\mcitedefaultmidpunct}
{\mcitedefaultendpunct}{\mcitedefaultseppunct}\relax
\EndOfBibitem
\bibitem{MGrabalosa:Thesis}
M.~Grabalosa, \emph{Flavour tagging developments within the LHCb experiment},
  CERN-THESIS-2012-075.\relax
\mciteBstWouldAddEndPunctfalse
\mciteSetBstMidEndSepPunct{\mcitedefaultmidpunct}
{}{\mcitedefaultseppunct}\relax
\EndOfBibitem
\bibitem{CDF:Dms:Thesis:2006}
N.~T.~Leonardo, \emph{Analysis of $B_s$ flavor oscillations at CDF},
  FERMILAB-THESIS-2006-18, 2006.\relax
\mciteBstWouldAddEndPunctfalse
\mciteSetBstMidEndSepPunct{\mcitedefaultmidpunct}
{}{\mcitedefaultseppunct}\relax
\EndOfBibitem
\bibitem{D0:Dms:Thesis:2008}
M.~S.~Anzelc, \emph{Study of mixing at the D{\O} detector at Fermilab using the
  semi-leptonic decay $B_s \to D_s\mu\nu{X}$}, FERMILAB-THESIS-2008-07,
  2008.\relax
\mciteBstWouldAddEndPunctfalse
\mciteSetBstMidEndSepPunct{\mcitedefaultmidpunct}
{}{\mcitedefaultseppunct}\relax
\EndOfBibitem
\bibitem{TBird:Thesis:2011}
T.~Bird, \emph{Towards measuring $B$ mixing in semileptonic decays at LHCb},
  CERN-THESIS-2011-184, 2011.\relax
\mciteBstWouldAddEndPunctfalse
\mciteSetBstMidEndSepPunct{\mcitedefaultmidpunct}
{}{\mcitedefaultseppunct}\relax
\EndOfBibitem
\bibitem{ROOT2}
R.~Brun and F.~Rademakers, \ifthenelse{\boolean{articletitles}}{{\it {\root -
  an object oriented data analysis framework}}, }{} vol.~389 of {\em {AIHENP'96
  Workshop, Lausanne}}, pp.~81--86, Sep, 1996.
\newblock
  doi:~\href{http://dx.doi.org/10.1016/S0168-9002(97)00048-X}{10.1016/S0168-9002(97)00048-X}\relax
\mciteBstWouldAddEndPuncttrue
\mciteSetBstMidEndSepPunct{\mcitedefaultmidpunct}
{\mcitedefaultendpunct}{\mcitedefaultseppunct}\relax
\EndOfBibitem
\bibitem{Verkerke:2003irmod}
W.~Verkerke and D.~Kirkby, \ifthenelse{\boolean{articletitles}}{{\it The
  \roofit toolkit for data modeling}, }{} in {\em 2003 Conference for Computing
  in High-Energy and Nuclear Physics ({CHEP} 03)}, ({La Jolla, California,
  USA}), March, 2003.
\newblock \href{http://arxiv.org/abs/physics/0306116}{{\tt
  arXiv:physics/0306116}}\relax
\mciteBstWouldAddEndPuncttrue
\mciteSetBstMidEndSepPunct{\mcitedefaultmidpunct}
{\mcitedefaultendpunct}{\mcitedefaultseppunct}\relax
\EndOfBibitem
\bibitem{LHCb-PAPER-2013-002}
LHCb collaboration, R.~Aaij {\em et~al.},
  \ifthenelse{\boolean{articletitles}}{{\it {Measurement of \CP-violation and
  the $B^0_s$-meson decay width difference with $B_s^0\to J/\psi K^+K^-$ and
  $B_s^0 \to J/\psi\pi^+\pi^-$ decays}},
  }{}\href{http://dx.doi.org/10.1103/PhysRevD.87.112010}{Phys.\ Rev.\  {\bf
  D87} (2013) 112010}, \href{http://arxiv.org/abs/1304.2600}{{\tt
  arXiv:1304.2600}}\relax
\mciteBstWouldAddEndPuncttrue
\mciteSetBstMidEndSepPunct{\mcitedefaultmidpunct}
{\mcitedefaultendpunct}{\mcitedefaultseppunct}\relax
\EndOfBibitem
\bibitem{Mike:GOF}
M.~Williams, \ifthenelse{\boolean{articletitles}}{{\it {How good are your fits?
  Unbinned multivariate goodness-of-fit tests in high energy physics}},
  }{}\href{http://dx.doi.org/10.1088/1748-0221/5/09/P09004}{JINST {\bf 5}
  (2010) P09004}, \href{http://arxiv.org/abs/1006.3019}{{\tt
  arXiv:1006.3019}}\relax
\mciteBstWouldAddEndPuncttrue
\mciteSetBstMidEndSepPunct{\mcitedefaultmidpunct}
{\mcitedefaultendpunct}{\mcitedefaultseppunct}\relax
\EndOfBibitem
\bibitem{Amoraal:2012qn}
J.~Amoraal {\em et~al.}, \ifthenelse{\boolean{articletitles}}{{\it {Application
  of vertex and mass constraints in track-based alignment}},
  }{}\href{http://dx.doi.org/10.1016/j.nima.2012.11.192}{Nucl.\ Instrum.\
  Meth.\  {\bf A712} (2013) 48}, \href{http://arxiv.org/abs/1207.4756}{{\tt
  arXiv:1207.4756}}\relax
\mciteBstWouldAddEndPuncttrue
\mciteSetBstMidEndSepPunct{\mcitedefaultmidpunct}
{\mcitedefaultendpunct}{\mcitedefaultseppunct}\relax
\EndOfBibitem
\bibitem{Fourier}
J.~B.~J. Fourier, \ifthenelse{\boolean{articletitles}}{{\it {Th\'eorie
  analytique de la chaleur}}, }{}Chez Firmin Didot, p\`ere et fils (1822)\relax
\mciteBstWouldAddEndPuncttrue
\mciteSetBstMidEndSepPunct{\mcitedefaultmidpunct}
{\mcitedefaultendpunct}{\mcitedefaultseppunct}\relax
\EndOfBibitem
\bibitem{FourierAnalysis}
S.~D. Conte and C.~de~Boor, {\em {Elementary numerical analysis}}, McGraw Hill
  Inc., 1980\relax
\mciteBstWouldAddEndPuncttrue
\mciteSetBstMidEndSepPunct{\mcitedefaultmidpunct}
{\mcitedefaultendpunct}{\mcitedefaultseppunct}\relax
\EndOfBibitem
\bibitem{Moser:1996xf}
H.~Moser and A.~Roussarie, \ifthenelse{\boolean{articletitles}}{{\it
  {Mathematical methods for $B^0$ anti-$B^0$ oscillation analyses}},
  }{}\href{http://dx.doi.org/10.1016/S0168-9002(96)00887-X}{Nucl.\ Instrum.\
  Meth.\  {\bf A384} (1997) 491}\relax
\mciteBstWouldAddEndPuncttrue
\mciteSetBstMidEndSepPunct{\mcitedefaultmidpunct}
{\mcitedefaultendpunct}{\mcitedefaultseppunct}\relax
\EndOfBibitem
\end{mcitethebibliography}
\ifx\mcitethebibliography\mciteundefinedmacro
\PackageError{LHCb.bst}{mciteplus.sty has not been loaded}
{This bibstyle requires the use of the mciteplus package.}\fi
\providecommand{\href}[2]{#2}

\end{document}